\documentclass[a4,12pt,reqno]{article}
\usepackage{amsmath}
\usepackage{amstext}
\usepackage{amsthm}
\usepackage{amssymb}
\usepackage[bottom]{footmisc}
\usepackage{upref}
\usepackage{cite}
\numberwithin{equation}{section}
\renewcommand{\baselinestretch}{1.57}
\begin{document}
\begin{titlepage} 
\renewcommand{\baselinestretch}{1.3}
\small\normalsize
\begin{flushright}
hep-th/0311196\\
MZ-TH/03-20
\end{flushright}

\vspace{0.1cm}

\begin{center}   

{\LARGE \textsc{renormalization group improved \\[2.35mm] gravitational actions:\\
[6.5mm]a brans-dicke approach}}

\vspace{1.4cm}
{\large M.~Reuter and H.~Weyer}\\

\vspace{0.7cm}
\noindent
\textit{Institute of Physics, University of Mainz\\
Staudingerweg 7, D-55099 Mainz, Germany}\\

\end{center}   
 
\vspace*{0.6cm}
\begin{abstract}
A new framework for exploiting information about the renormalization group
(RG) behavior of gravity in a dynamical context is discussed. The 
Einstein-Hilbert action is RG-improved by replacing Newton's constant and the
cosmological constant by scalar functions in the corresponding Lagrangian
density. The position dependence of $G$ and $\Lambda$ is governed by a RG
equation together with an appropriate identification of RG scales with points
in spacetime. The dynamics of the fields $G$ and $\Lambda$
does not admit a Lagrangian description in general. Within the Lagrangian
formalism for the gravitational field they have the status of externally 
prescribed ``background'' fields. The metric satisfies an effective Einstein 
equation similar to that of Brans-Dicke theory. Its 
consistency imposes severe constraints on allowed backgrounds. In the new
RG-framework, $G$ and $\Lambda$ carry energy and momentum. It is tested in the
setting of homogeneous-isotropic cosmology and  is compared to alternative
approaches where the fields $G$ and $\Lambda$ do not carry gravitating
4-momentum. The fixed point regime of the underlying RG flow is studied
in detail. 

04.50.+h, 04.60.-m, 11.10.Hi
\end{abstract} 
\end{titlepage}
%
%
\section{Introduction}
It is an old idea that Newton's ``constant'' actually is not really constant
throughout spacetime but varies from one spacetime point to another. The
perhaps most popular self-consistent theory of gravity which implements this
idea is Brans-Dicke theory \cite{bradi}. Originally devised in an attempt
at modifying General Relativity so as to become compatible with 
Mach's principle, it is by now the prototype of a theory in which the 
gravitational interaction is mediated by the metric together with some
additional non-geometric field, here a scalar $\phi$. The Brans-Dicke
field $\phi$ is introduced as the inverse of the position dependent
Newton constant: $\phi \left( x \right) \equiv 1 / G \left( x \right)$.
In the original form of the theory its dynamics, along with that of the
metric, is derived from the action
\begin{align}
S_{\text{\scriptsize BD}} & =
\frac{1}{16 \pi} \, \int \! \! \text{d}^{4} x~
\sqrt{-g} ¸\, \left( \phi \, R - \omega \, \phi^{-1} \,
\partial_{\mu} \phi \, \partial^{\mu} \phi \right) 
+ S_{\text{M}}. \label{1.1}
\end{align}
It supplements the Einstein-Hilbert term with an (almost) conventional scalar
kinetic term. ($S_{\text{M}}$ denotes the matter action.) Varying
$S_{\text{\scriptsize BD}}$ with respect to the
metric $g_{\mu \nu}$ and $\phi$ leads, respectively, to a modified
Einstein equation,
\begin{align}
G_{\mu \nu} & = 
8 \pi \, \phi^{-1} \, \left( T_{\mu \nu} + {\mathcal T}_{\mu \nu}^{\omega}
 \right)
+ \phi^{-1} \, \left( D_{\mu} D_{\nu} \phi
- g_{\mu \nu} \, D^{2} \phi \right), \label{1.2}
\end{align}
and the scalar equation of motion
\begin{align}
 D^{2} \phi & = 
\frac{8 \pi}{3+ 2 \omega} \, T_{\mu}^{~\mu}. \label{1.3}
\end{align}
Here $T_{\mu \nu}$ is the energy-momentum tensor obtained from $S_{\text{M}}$,
and
\begin{align}
{\mathcal T}_{\mu \nu}^{\omega} & = 
\frac{\omega}{8 \pi \, \phi} \,
\Bigl[ D_{\mu} \phi \, D_{\nu} \phi
- \tfrac{1}{2} \, g_{\mu \nu} \,
D_{\rho} \phi \, D^{\rho} \phi \Bigr] \label{1.4}
\end{align}
is the one stemming from the kinetic term of $\phi$; its normalization
$\omega$ is a free parameter a priori. Together with the 
$\phi^{-1} \, D D \phi$-terms in \eqref{1.2}, originating
from the $x$-dependence of $\phi$ in the Lagrangian $\sqrt{-g} \, \phi
\, R$, the tensor ${\mathcal T}_{\mu \nu}^{\omega}$ describes the 4-momentum
carried by the scalar field $\phi$. The point to be emphasized here is that in
Brans-Dicke theory the $x$-dependence of Newton's constant is governed by a 
simple local equation of motion with an obvious Lagrangian or
Hamiltonian formulation, the Klein-Gordon equation \eqref{1.3}.

Recently a lot of work was devoted to a different type of theories with
a variable Newton constant where the dynamics of $G \left( x \right)$ does not
admit a straightforward Lagrangian description and simple local equations
of motion such as \eqref{1.3}.
These theories arise by ``renormalization group (RG) improving'' classical
General Relativity, i.\,e.\ by replacing $G$ and similar generalized couplings
such as the cosmological constant $\Lambda$ by scale dependent or ``running''
quantities \cite{bertodm}. Here the starting point is a scale dependent
effective action for the gravitational field, $\Gamma_{k} \left[ g_{\mu \nu}
\right]$, a Wilson-type (``coarse-grained'') free energy functional. It defines
an effective field theory valid near the mass scale $k$ or length scale
$\ell \equiv k^{-1}$. This means that in order to include all fluctuation effects
relevant at $k$ it is sufficient to employ $\Gamma_{k}$ at tree level. In the
context of quantum field theory, $\Gamma_{k}$ obtains
 from the fundamental
(bare) action of the theory by integrating out all quantum fluctuations with 
momenta larger than the infrared cutoff $k$, i.\,e.\ wavelengths smaller than
$\ell$. The ``effective average action'' \cite{avact,avactrev} 
is a concrete 
realization of this idea; following similar lines as in Yang-Mills theory
\cite{ym} it has been applied to the quantized gravitational field
\cite{mr} and to quantum gravity with matter \cite{percadou}. Another
logical possibility is that the scale-dependence or ``running'' of 
$\Gamma_{k}$ arises as the result of a purely classical averaging process
\cite{carfora}.

In either case the $k$-dependence of $\Gamma_{k}$ is governed by a functional
differential equation, the ``flow equation'' or ``exact RG equation''
\cite{bagber}. A general functional $\Gamma_{k} \left[ g_{\mu \nu}
\right]$ can be parameterized by infinitely many dimensionless couplings related
to the coefficients of the field monomials (higher powers of the curvature,
non-local terms \cite{cwnonloc}, etc.\ ). They include the dimensionless Newton
constant $g \left( k \right) \equiv k^{2} \, G \left( k \right)$ and 
cosmological constant $\lambda \left( k \right) \equiv  
\Lambda \left( k \right) / k^{2}$. 
When expressed in terms of the running coupling
constants the flow equation assumes the form of a system of infinitely many 
ordinary coupled differential equations:
\begin{align}
\begin{split}
k \, \frac{\text{d}}{\text{d} k} \, \lambda \left( k \right)
& = \boldsymbol{\beta}_{\lambda} \left( \lambda, g, \cdots \right) \\
k \, \frac{\text{d}}{\text{d} k} \, g \left( k \right)
& = \boldsymbol{\beta}_{g} \left( \lambda, g, \cdots \right) \\
& \vdots
\end{split}
\label{1.5}
\end{align}
A solution $k \mapsto \left( \lambda \left( k \right), 
g \left( k \right), \cdots \right)$ corresponds to a RG trajectory on
``theory space'', the space of all action functionals. Because of the 
technical complexity of the problem one is often forced to restrict the RG
flow to a finite-dimensional subspace, a truncated theory space. In the
``Einstein-Hilbert truncation'', say, only the couplings $g$ and $\lambda$ are
taken into account; every solution $\left( \lambda \left( k \right), 
g \left( k \right) \right)$ of the truncated flow equation corresponds
to the one-parameter family of action functionals
\begin{gather}
\Gamma_{k} \left[ g_{\mu \nu} \right] =
\frac{1}{16 \pi G \left( k \right)} \,
\int \! \! \text{d}^{4} x ~ \sqrt{-g} \,
\bigl[ R - 2 \, \Lambda  \left( k \right) \bigr] \label{1.6}
\intertext{with $G$ and $\Lambda$ given by}
G \left( k \right) = g \left( k \right) / k^{2} 
\quad \text{and} \quad
\Lambda \left( k \right) = \lambda \left( k \right) \, k^{2}. \label{1.7}
\end{gather}
Accidently, in this truncation the resulting effective field equations ( with
a matter term added) look like the standard Einstein equation:
\begin{gather}
G_{\mu \nu} = - \Lambda \left( k \right) \, g_{\mu \nu} + 8 \pi 
G \left( k \right) \, T_{\mu \nu}. \label{1.7'}
\end{gather}
For Quantum Einstein Gravity, truncated flow equations were derived in 
\cite{mr}, \cite{percadou}, and \cite{oliver1} - \cite{frank2}, for
instance. In ref.\ \cite{frank2} a nonlocal truncation ansatz had been
used \cite{venez}.

Once the $k$-dependence of $G$, $\Lambda$, and the other parameters 
is known for the
RG trajectory of interest one can try to use this information in order
to ``RG improve'' the predictions of standard General Relativity.
One looks for a ``cutoff identification'' of the form \cite{nelson}
\begin{gather}
k = k \left( x \right) \label{1.8}
\intertext{which would convert the scale-dependence of $G$, $\Lambda$, $\cdots$
to a position-dependence:}
G \left( x \right) =  G \left( k \left( x \right) \right), \quad
\Lambda \left( x \right) =  \Lambda \left( k \left( x \right) \right), \quad
\cdots \label{1.9}
\end{gather}
It is certainly not always possible to associate cutoff values to spacetime
points and to find an appropriate function $k \left( x \right)$. However, in
physical situations with a high degree of symmetry, very often symmetry 
arguments and dimensional analysis lead to essentially unique answers for
$k \left( x \right)$. Let us mention two examples where a natural and
physically transparent cutoff identification suggests itself.

We consider the cosmology of a homogeneous and isotropic Universe and
assume that the metric has been brought to the standard Robertson-Walker form
with a scale factor $a \left( t \right)$. Then the postulate of homogeneity
and isotropy implies that $k$ can depend on the cosmological time only, either
explicitly or implicitly via the scale factor: 
$k = k \left( t, a \left( t \right), \dot a \left( t \right), \cdots \right)$.
In ref.\ \cite{cosmo1} we gave detailed arguments as to why the purely explicit
time dependence
\begin{gather}
k \left( t \right) = \xi / t \label{1.10}
\end{gather}
with $\xi$ a positive constant of order unity is the correct identification
in a first approximation. The reason is that, when the age of the Universe is
$t$, clearly no quantum or classical fluctuation with a frequency smaller than
$1/t$ can have played any role yet. Hence the integrating-out of modes
or ``coarse-graining'' which underlies the Wilson renormalization group should
be stopped at $k \approx 1/t$.  Moreover, for many cosmologies of interest
other plausible cutoffs such as $k=H \equiv \dot a / a$ are equivalent 
to \eqref{1.10}. (The cutoff identification \eqref{1.10} has been used in
refs.\ \cite{cosmo1,cosmo2,esposito} in order to improve the
cosmological evolution \textit{equations} of General Relativity.)

In ref.\ \cite{bh} the RG-improvement of a Schwarzschild black hole,
i.\,e.\ of a \textit{solution} to the standard Einstein equation has been
discussed. Here the symmetries imply that $k$ can be dependent on the
Schwarzschild radial coordinate $r$ only. The natural cutoff identification
which can be motivated in various ways \cite{bh} is
\begin{gather}
k \left( r \right) = \xi / d \left( r  \right). \label{1.11}
\end{gather}
Here $d \left( r \right)$ is the proper distance from a point with 
coordinate $r$ to the center of the black hole. In \cite{bh}, $d \left(
r \right)$ was computed from the unperturbed Schwarzschild metric;
in a more refined treatment the improved metric should be used instead.

Let us now discuss the physical cutoff mechanism in the generic
case where the problem is not highly symmetric with a single preferred
scale. Within the effective average action formalism the general rule
is to follow the RG flow from the bare action
$\Gamma_{k=\infty} \equiv S$, which serves as the initial condition
for the $\Gamma_{k}$-flow equation, all the way down to $k=0$. The endpoint
of the RG trajectory thus obtained, $\Gamma_{k=0} \equiv \Gamma$,
is nothing but the ordinary effective action \cite{avactrev}.
If we have no a priori knowledge about a physical cutoff, what we have to do 
is to solve the effective equation of motion
\begin{align}
\frac{\delta \Gamma \bigl[ g_{\mu \nu} \bigr]}{\delta g_{\mu \nu}} =0
\label{E.1}.
\end{align}
It is well-known that, despite its classical appearance, this equation
is fully quantum mechanical, and metrics $g_{\mu \nu} \equiv
<\!\!g_{\mu \nu}\!\!>$ satisfying it are exact quantum vacuum solutions.

In practice, in any realistic theory, it is impossible to compute
the RG trajectories exactly. One is forced to truncate the theory space, and 
very often the truncations used are reliable only for large values of $k$
(UV) but not for small ones (IR). Typically, and in particular for asymptotically
free theories, simple local truncations are sufficient in the UV, but for
$k \to 0$ nonlocal terms must be included in the truncation ansatz for
$\Gamma_{k}$. In particular in massless theories it is technically extremely
difficult to handle those nonlocal terms. In QCD, say, a reliable calculation
of $\Gamma_{k}$ for $k \to 0$ is still out of reach, and the corresponding
problems in Quantum Gravity are much harder even \cite{frank1,frank2}. While
$\Gamma_{k \to 0}$ is not available in full generality, the method outlined
above (inserting $k = k \left( x \right)$ into a fairly simple truncated
RG trajectory) is a kind of ``short-cut'' for the way from the UV to the IR.
Even though we use a strictly local truncation, the cutoff identification
$k = k \left( x \right)$ introduces nonlocal features into the theory which,
under certain conditions, are equivalent to some of the elusive nonlocal
terms in $\Gamma_{k}$ or the $\Gamma$ of the standard approach. In order to
explain this partial equivalence we must review the phenomenon of
\textit{decoupling} in the formalism of the effective average action
\cite{avact,avactrev}.

For the sake of simplicity let us discuss the universality class of
a single real, $Z_{2}$-symmetric scalar field $\Phi \left( x \right)$
in flat Euclidean space. Its effective average action 
$\Gamma_{k} \left[ \Phi \right]$ is to be determined from the flow
equation \cite{avact}
\begin{align}
k \, \tfrac{\text{d}}{\text{d} k} \, \Gamma_{k} \left[ \Phi \right]
& =
\tfrac{1}{2} \, \text{Tr} \, \Bigl[ \bigl( \Gamma_{k}^{(2)} 
+ R_{k} \bigr)^{-1} \, k \, \tfrac{\text{d}}{\text{d} k} \, R_{k}
\Bigr]
\label{E.2}
\end{align}
subject to the initial condition $\Gamma_{\infty} =S$. Here
$\Gamma_{k}^{(2)}$ denotes the infinite dimensional matrix of second
functional derivatives of $\Gamma_{k} \left[ \Phi \right]$, and $R_{k}$ is 
the cutoff function. For the purposes of the present argument it is sufficient
to employ a $R_{k}$ of ``mass type'': $R_{k} = k^{2}$ \cite{avactrev}.
An important class of nonperturbative solutions to \eqref{E.2} can be found
with the ansatz
\begin{align}
\Gamma_{k} \left[ \Phi \right] = \int \text{d}^{4} x~
\Big \{ \tfrac{1}{2} Z \left( k \right) \, \partial_{\mu} \Phi
\, \partial^{\mu} \Phi + \tfrac{1}{2} m^{2} \left( k \right) \, \Phi^{2}
+ \tfrac{1}{12} \, \lambda \left( k \right) \, \Phi^{4} + \cdots \Big \}.
\label{E.3}
\end{align}
To start with, we neglect the running of the kinetic term (``local potential
ansatz'') and approximate $Z (k) \equiv 1$. For functionals of this type, and
in a momentum basis where $- \partial^{2} \overset{\wedge}{=} p^{2}$,
the denominator appearing under the trace of \eqref{E.2} reads
\begin{align}
\Gamma_{k}^{(2)} + R_{k} & =
p^{2} + m^{2} (k) + k^{2} + \lambda (k) \, \Phi^{2} + \cdots.
\label{E.4}
\end{align}
In a diagrammatic loop calculation of $\Gamma_{k}$ it is the inverse of
\eqref{E.4}, evaluated at $\Phi = <\!\! \Phi \!\!>$, which appears as the 
effective propagator in all loops. It contains an IR cutoff at the scale $k$,
a simple mass term $k^{2}$ which adds to $m^{2} (k)$ in the special case 
considered here. (In general $R_{k} \equiv R_{k} (p^{2})$ introduces a
$p^{2}$-dependent mass.)
The $p_{\mu}$-modes (plane waves) are integrated out efficiently only in the
domain $p^{2} \gtrsim m^{2} + k^{2} + \lambda \, \Phi^{2} + \cdots$.
In the opposite case all loop contributions are suppressed by the 
effective mass square $m^{2} + k^{2} + \lambda \, \Phi^{2} + \cdots$.
It is the sum of the ``artificial'' cutoff $k^{2}$, introduced in order to
effect the coarse graining, and the ``physical'' cutoff terms 
$m^{2} (k) + \lambda (k) \, \Phi^{2} + \cdots$. As a consequence,
$\Gamma_{k}$ displays a significant dependence on $k$ only if
$k^{2} \gtrsim m^{2} (k) + \lambda (k) \, \Phi^{2} + \cdots$ because otherwise
$k^{2}$ is negligible relative to $m^{2} + \lambda \, \Phi^{2} + \cdots$
in all propagators; it is then the physical cutoff scale
$m^{2} + \lambda \, \Phi^{2} + \cdots$ which delimits the range of 
$p^{2}$-values which are integrated out. Typically, for $k$ very large,
$k^{2}$ is larger than the physical cutoffs so that $\Gamma_{k}$ ``runs''
very fast.
Lowering $k$ it might happen that, at some $k=k_{\text{dec}}$, the 
``artificial'' cutoff $k$ becomes smaller than the running mass $m(k)$. At
this point the physical mass starts playing the role of the actual cutoff;
its effect overrides that of $k$ so that $\Gamma_{k}$ becomes
approximately independent of $k$ for $k < k_{\text{dec}}$.
As a result, $\Gamma_{k} \approx \Gamma_{k_{\text{dec}}}$ for all $k$ below
the threshold $k_{\text{dec}}$, and in particular the ordinary effective
action $\Gamma = \Gamma_{0}$ does not differ from $\Gamma_{k_{\text{dec}}}$
significantly. This is the prototype of a ``decoupling'' or ``threshold''
phenomenon.

The situation is more interesting when $m^{2}$ is negligible and
  $k^{2}$ competes with $\lambda \, \Phi^{2}$ for the role of the actual cutoff.
(Here we assume that $\Phi$ is $x$-independent.)
The running of $\Gamma_{k}$, evaluated at a fixed $\Phi$, stops once
$k \lesssim k_{\text{dec}} (\Phi)$ where the by now field dependent decoupling
scale obtains from the implicit equation 
$k_{\text{dec}}^{2} = \lambda \bigl( k_{\text{dec}} \bigr) \, \Phi^{2}$.
Decoupling occurs for sufficiently large values of $\Phi$, the RG evolution
below $k_{\text{dec}}$ is negligible then; hence, at $k=0$,
\begin{align}
\Gamma \left[ \Phi \right] & =
\Gamma_{k} \left[ \Phi \right] \Big \rvert_{k=k_{\text{dec}} (\Phi)}.
\label{E.5}
\end{align}
Eq.\ \eqref{E.5} is an extremely useful tool for effectively going beyond the
truncation \eqref{E.3} without having to derive and solve a more complicated
flow equation. In fact, thanks to the additional $\Phi$-dependence which comes
into play via $k_{\text{dec}} (\Phi)$, eq.\ \eqref{E.5} can predict certain
terms which are contained in $\Gamma$ even though they are not present in the 
truncation ansatz.

A simple example illustrates this point. For $k$ large,
the truncation \eqref{E.3} yields a logarithmic running of the
$\Phi^{4}$-coupling: $\lambda (k) \propto \ln (k)$.
As a result, \eqref{E.5} suggests that $\Gamma$ should contain a term
$\propto \ln \bigl( k_{\text{dec}} (\Phi) \bigr) \, \Phi^{4}$.
Since, in leading order, $k_{\text{dec}} \propto \Phi$, this leads us to
the prediction of a $\Phi^{4} \, \ln (\Phi)$-term in the conventional
effective action. This prediction, including the prefactor of the term, is
known to be correct actually: the Coleman-Weinberg potential of massless
$\Phi^{4}$-theory does indeed contain a $\Phi^{4} \, \ln (\Phi)$-term.
Note that this term admits no power series expansion in $\Phi$, so it
lies outside the space of functionals spanned by the original ansatz
\eqref{E.3}.

This example nicely illustrates the power of the decoupling arguments.
They can be applied even when $\Phi$ is taken $x$-dependent as it is
necessary for computing $n$-point functions by differentiating
$\Gamma_{k} \left[ \Phi \right]$. The running inverse propagator is given
by $\Gamma_{k}^{(2)} (x-y) = \delta^{2} \Gamma_{k} / \delta \Phi (x) \,
\delta \Phi (y)$, for example. Here a new potential cutoff scale enters the
game: the momentum $q$ dual to the distance $x-y$. When it serves as the acting
IR cutoff, the running of $\widetilde \Gamma_{k}^{(2)} (q)$, the Fourier 
transform of $\Gamma_{k}^{(2)} (x-y)$, stops once $k^{2}$ is smaller than
$k_{\text{dec}}^{2} = q^{2}$. Hence $\widetilde \Gamma_{k}^{(2)} (q)
\approx \widetilde \Gamma_{k}^{(2)} (q) \Big \rvert_{k = \sqrt{q^{2}\,}}$
for $k^{2} \lesssim q^{2}$, provided no other physical scales intervene.
As a result, if one allows for a running $Z$-factor in the truncation \eqref{E.3}
one predicts a propagator of the type $\left[ Z \left( \sqrt{q^{2}\,} \right) 
\, q^{2} \right]^{-1}$ in the standard effective action. Note that it
corresponds to a nonlocal term 
$\propto \Phi \, Z \left( \sqrt{- \partial^{2}\,} \right) \, \partial^{2}
\Phi$ in $\Gamma$, even though the truncation ansatz was perfectly local.

At this point also the origin of the $x$-dependent scale $k=k(x)$
employed in the present paper becomes clear: The distance 
$|x-y| \equiv r$ translates to a momentum $\sqrt{q^{2}\,} \approx 1/r$
so that the running of $Z (k)$ is stopped at the physical scale $1/r$.
This is precisely the interpretation we use in Quantum Gravity, with
$\sqrt{g\,} \, R / G (k)$ taking the place of $Z (k) \, \bigl(
\partial_{\mu} \Phi \bigr)^{2}$, and \eqref{1.11} covariantizing
$k_{\text{dec}} \propto 1/r$.

More generally, by inserting $k=k(x)$ into $G (k)$ we try to mimic the
effect of terms contained in $\Gamma$, but not in the Einstein-Hilbert 
truncation. In this sense the strategy is similar to the derivation of the
$\Phi^{4} \, \ln (\Phi)$- and the nonlocal kinetic term above.

RG improvement based upon the above ideas on decoupling is a powerful
tool, a kind of ``short-cut'' from the UV to the IR, if one is able to 
identify the actual physical cutoff without solving for the full RG flow. This
requires at detailed case-by-case study of the physical problem at hand;
for cosmology and black holes see refs.\ \cite{cosmo1} and \cite{bh} for
a justification of \eqref{1.10} and \eqref{1.11}, respectively.
The identification of the physical cutoff is greatly facilitated by
symmetry arguments and, in particular near a scale invariant fixed point, by
dimensional analysis. But even in the generic case it is sometimes possible to
get a handle on the essential physics by a careful analysis of the various 
potential cutoff terms in $\Gamma_{k}^{(2)}$.

Let us return to gravity now.
Generally speaking, every pair consisting of a RG trajectory
$\left( \lambda \left( k \right), g \left( k \right), \cdots \right)$ and a 
cutoff identification $k = k \left( x  \right)$ generates a set of 
position dependent ``constants'' $\Lambda \left( x \right)$, 
$G \left( x \right)$, $\cdots$. Clearly the dynamical origin of the
$x$-dependence of $G$, say, is rather different from what happens in
Brans-Dicke theory. The basic ``equation of motion'' is the exact RG equation.
It is to be solved before $k \left( x \right)$ is inserted and has no 
connection to any specific spacetime therefore, in contradistinction
to the Klein-Gordon equation for $\phi$. A closely related fact, which will
become very important later on and is in fact one of the main motivations for
the present paper, is the following: 
In general there exists no Lagrangian formulation
of the system ``RG equation plus cutoff identification'' which could take the
place of the $\phi^{-1} \, \partial_{\mu} \phi \, \partial^{\mu}
\phi$-term.\footnote{Of course the exact framework (construct $\Gamma$, solve
\eqref{E.1}) is Lagrangian, but this is of no help if we take the 
``short-cut'' discussed above.}
While one may continue to specify the gravitational dynamics by means of a 
Einstein-Hilbert-like action, the couplings $\Lambda \left( x \right)$, 
$G \left( x \right)$, $\cdots$ have the status of externally prescribed fields.
In a sense, we are dealing here with a kind of background field problem:
the metric $g_{\mu \nu}$ must be determined in presence of the
``background fields'' $\left( \Lambda \left( x \right), G \left( x \right), 
\cdots \right)$ on which it functionally depends therefore. We shall see that
not all backgrounds are admissible and it will be an important
question which ones lead to consistent field equations for $g_{\mu \nu}$.

In the rest of this paper we restrict the discussion to the Einstein-Hilbert
truncation. Let us assume we have solved its RG equation and obtained
 $\left( \Lambda \left( x \right), G \left( x \right) \right)$ from some
RG trajectory. How can we take advantage of this information? A possible
strategy is to insert the identification $k = k \left( x \right)$ into the
effective field equation \eqref{1.7'},
\begin{gather}
G_{\mu \nu} = - \Lambda \left( x \right) \, g_{\mu \nu} + 8 \pi 
G \left( x \right) \, T_{\mu \nu}, \label{1.12}
\end{gather}
and to solve this differential equation in presence of the scalar 
background fields $G$ and $\Lambda$. In refs.\ \cite{cosmo1,cosmo2,kalligas} 
this approach was employed in the context of cosmology.
Some of the corresponding results which we shall need later on are collected
in Appendix \ref{appendix A}.

This strategy, henceforth referred to as the approach of 
\textit{improving equations}, is not always meaningful. For instance, if we are
interested in the structure of black holes in absence of matter 
($T_{\mu \nu} =0$) and at scales where the cosmological constant can be
neglected, the improved field equation \eqref{1.12} reduces to 
$G_{\mu \nu}=0$. This is the standard vacuum Einstein equation which does
not know anything about the RG running. Here the leading corrections can be
taken into account by \textit{improving solutions} of the classical Einstein
equation. The Schwarzschild solution, for instance, satisfies $G_{\mu \nu}=0$
but it contains $G$ as a constant of integration. The improved Schwarzschild
metric is obtained by replacing this constant with $G \left( x \right)
\equiv G \left( r \right)$. (See ref.\ \cite{bh} for a detailed discussion of
this metric.)

The approach of improving solutions is much less powerful than that of 
improving equations. The former is reliable only if the original and improved
metrics are not very different, while the latter might well lead to solutions
of the improved equations which are quite different from the corresponding
classical ones, without necessarily being unreliable \cite{cosmo1}.

The limited applicability of the ``improving solutions'' method is one of our
motivations for trying to find a new way of injecting the information
provided by the renormalization group into the gravitational field equations.
The approach we are going to investigate in the present paper is that of
\textit{improving actions}. The basic idea is to make the cutoff identification
in the action functional, i.\,e.\ before the derivation of the field
equations. We start from the $k$-dependent Einstein-Hilbert action \eqref{1.6}
and substitute $k=k \left( x \right)$ in the corresponding Lagrangian density:
\begin{align}
S_{\text{mEH}} \left[ g,G,\Lambda \right]& =  
\frac{1}{16 \pi} \int \!\! \text{d}^{4} x~
\sqrt{-g} \, \biggl \{ \frac{R}{G \left( x \right)} -
2 \, \frac{\Lambda \left( x \right)}{G \left( x \right)} \biggr \}.
\label{2.1}
\end{align}
When varied with respect to $g_{\mu \nu}$, this modified Einstein-Hilbert
(mEH) functional gives rise to field equations which differ from those of the
``improving equations'' approach by terms involving derivatives of
$G \left( x \right)$. The action $S_{\text{mEH}}$ is not supposed to be
varied with respect to $G$ and $\Lambda$. The functions $G \left( x \right)$
and $\Lambda \left( x \right)$ are obtained from a RG trajectory in the way
described above, hence are external to the Lagrangian formalism which we use
for the gravitational field. In this sense we are dealing with
an ``external field Brans-Dicke theory'' where the 
$g_{\mu \nu}$-dynamics takes place in the fixed background of the scalar
fields $G \left( x \right)$ and $\Lambda \left( x \right)$.

Not all backgrounds lead to a consistent equation for the metric. Both when
improving the field equation and the action the resulting modified Einstein
equation has the form $G_{\mu \nu} = {\mathcal E}_{\mu \nu}$ where
${\mathcal E}_{\mu \nu}$ is constructed from $T_{\mu\nu}$, $G$, $\Lambda$,
and their derivatives.
Since $D^{\mu} G_{\mu \nu} =0$ by Bianchi's
identity, consistency requires that $D^{\mu} {\mathcal E}_{\mu \nu} =0$.
In classical General Relativity this condition is satisfied if $T_{\mu \nu}$
is conserved. In the present case additional conditions on
$G \left( x \right)$ and $\Lambda \left( x \right)$ arise, henceforth referred to
as ``consistency conditions''.

Only if the background satisfies the consistency condition Einstein's equation
can be integrated. However, it should be remarked that in practice this
restriction very often is not a drawback but rather an advantage of the formalism. The 
point to be remembered here is that most features of the RG trajectories
are unphysical, i.\,e.\ not directly observable, and can be changed by changing
the cutoff scheme. The effective average action, for example, has a built-in
``shape function'' which controls the transition from the momenta which are
integrated out to those which are not \cite{mr,oliver1}. Changing the
shape function changes the trajectories. Only quantities which, in the language
of statistical mechanics, are ``universal'' remain invariant and thus qualify
for the status of an observable. Therefore, when one realizes that some
background does not satisfy the consistency conditions one can try to change
the cutoff scheme (shape function) and/or the identification $k = k \left(
x \right)$ so as to achieve consistency. If we are successful, there is a 
non-trivial conspiracy between the RG equation and the modified Einstein
equation which can teach us something about the correct mathematical model
(shape function, cutoff identification) of the physical mechanism which
stops the RG running at some scale. (See refs.\ \cite{bh,cosmo1}
for a detailed discussion of this point.)

Finally let us say a few words about the RG trajectories which we are going to
employ in the present paper. They are motivated by the explicit results 
obtained from the truncated flow equation of 
Quantum Einstein Gravity \cite{mr,oliver1,frank1,oliver2,percacciperini,frank2,souma}, and by a
phenomenologically inspired conjecture formulated in \cite{cosmo2}.

Within the Einstein-Hilbert truncation of the effective average action,
 the RG equations for 
$g \left( k \right)$ and $\lambda \left( k \right)$ were first derived in
\cite{mr} and solved numerically in \cite{frank1}. The RG flow on the 
$g$-$\lambda$-plane is dominated by two fixed points 
$\left( g_{*}, \lambda_{*} \right)$: a Gaussian fixed point at
$g_{*}= \lambda_{*}=0$, and a non-Gaussian one with $g_{*}>0$ and
$\lambda_{*}>0$ \cite{souma}. The high-energy or short-distance behavior of 
Quantum Einstein Gravity is governed by the non-Gaussian fixed point:
for $k \to \infty$, all RG trajectories run into this fixed point. If it is
present in the exact flow equation, it can be used to construct a fundamental,
i.\,e.\ microscopic quantum theory of gravity by taking the limit of infinite
UV cutoff along one of the trajectories running into the fixed point. This 
corresponds precisely to Weinberg's asymptotic safety scenario \cite{wein}:
performing the UV limit at a fixed point one can be sure that the theory does
not develop uncontrolled singularities at high energies. 

In refs.\
\cite{oliver1,oliver2,percacciperini} detailed consistency checks were 
performed which indicate that the non-Gaussian fixed point should be a 
reliable prediction of the Einstein-Hilbert truncation, and that Quantum
Einstein Gravity has very good chances of being a non-perturbatively
renormalizable (and not only an effective) field theory of gravity. A 
conceptually independent investigation which points in exactly the same
direction is the quantization of the 2 Killing vector-reduction of Einstein
gravity in ref.\ \cite{max}.

At the fixed point, the dimensionless couplings assume constant values
$g_{*}$ and $\lambda_{*}$, respectively. Hence the dimensionful ones run
according to
\begin{gather}
G \left( k \right) =  g_{*} / k^{2}, \quad
\Lambda \left( k \right) =  \lambda_{*} \, k^{2} \label{1.13}.
\end{gather}
In particular, $G \left( k \right) \to 0$ for $k \to \infty$, so that
Newton's constant is an asymptotically free coupling.

At the other end of the energy scale, for $k \to 0$, the trajectory with 
vanishing cosmological constant hits the Gaussian fixed point. This is the
realm of classical General Relativity where $G \left( k \right)$ assumes
an approximately $k$-independent, non-zero value. In this ``perturbative
regime'' \cite{oliver1} the leading quantum corrections can be computed
as a power series in $k/m_{\text{Pl}}$ where 
$m_{\text{Pl}} \equiv \left[G \left( k=0 \right) \right]^{-1/2}$ is the
Planck mass \cite{mr,oliver1}. For intermediate values of $k$, in particular
at the cross-over from the UV- to the IR-fixed point, the flow equations must
be solved numerically \cite{frank1}.

In this paper we are going to RG-improve homogeneous-isotropic cosmologies
for which \eqref{1.10} is the appropriate cutoff identification. Given
the (numerical) solutions for $G \left( k \right)$ and $\Lambda \left( k \right)$
we may replace $k$ with $\xi /t$ and obtain the corresponding ``background''
fields $G \left( t  \right)$ and $\Lambda \left( t \right)$. In particular, 
near a fixed point, we have
\begin{subequations} \label{1.50}
\begin{align}
G \left( t \right) & = \widetilde g_{*} \, t^{2} \label{1.50a} \\
\Lambda \left( t \right) & = \widetilde \lambda_{*} \, t^{-2} \label{1.50b}
\end{align}
\end{subequations}
with the constants
\begin{subequations} \label{1.51}
\begin{align}
\widetilde g_{*} & \equiv  g_{*} \, \xi^{-2} \label{1.51a} \\
 \widetilde \lambda_{*} & \equiv \lambda_{*} \, \xi^{2}. \label{1.51b}
\end{align}
\end{subequations}
Since the time dependence \eqref{1.50} applies in the vicinity of a UV
fixed point it is realized in the very \textit{early} Universe,
i.\,e.\ for $t \to 0$.

In \cite{cosmo2} it has been conjectured that in the asymptotically 
\textit{late} Universe, too, $G$ and $\Lambda$ have a time dependence given
by \eqref{1.50}, with different values of $g_{*}$ and $\lambda_{*}$ though.
According to this conjecture, the $t \to \infty$ behavior of the Universe
is governed by a further non-Gaussian fixed point. It does not occur
within the Einstein-Hilbert truncation but there are first indications 
\cite{frank2,venez} that the inclusion of non-local invariants could have
the desired effect. The main motivation for this conjecture is that it explains
the approximate equality of vacuum and matter energy density in the present
Universe in a very elegant and natural way \cite{cosmo2}.

The purpose of the present paper is twofold. In the first part we discuss the
general theory of RG-improved gravitational actions, and in the second we test
and illustrate this approach in the context of cosmology\footnote{For related 
work within standard Brans-Dicke theory see \cite{bertocosmo} and references
therein.}. To be as explicit
as possible and to obtain analytic results, most of the time we shall use
the ``fixed point background'' \eqref{1.50} as an example\footnote{Using a 
different approach, RG-improved cosmologies
with other time dependencies of $\Lambda$
were investigated in \cite{sola}. A general discussion of cosmologies
with a time dependent $\Lambda$ can be found in \cite{mrcwcosmo}.}. This allows us to
compare the results obtained by ``improving actions'' with what had been found
by ``improving equations''. (The latter results are briefly summarized in
Appendix \ref{appendix A}.)

The remaining sections of this paper are organized as follows. In Section
\ref{3} we develop the general framework describing the gravitational dynamics
in the background of a position-dependent $G$ and $\Lambda$ which results 
from RG improving action functionals of the Einstein-Hilbert type. In
particular we identify various classes of solutions to the equations of 
motion (``Class I, II, III'') which enjoy special properties; some of them
can be obtained in a very efficient way by means of a Weyl transformation.
In Section \ref{4} the dynamical equations and consistency conditions
are specialized for the cosmology of homogeneous and isotropic Universes.
Then, in Sections \ref{5} - \ref{7}, we find solutions to the cosmological
evolution equations belonging to Class I, II, and III, respectively.
They illustrate a number of general properties of our approach in a 
particularly transparent way. In particular they clarify its relation to
alternative strategies for the RG improvement of gravity (improvement at
the level of the field equations). The results are summarized in Section
\ref{8}.

Many of the cosmological solutions found in this paper are not yet
realistic from the physical point of view. Their
importance resides in the fact that they provide us with valuable insights
into the general features of the improved action-approach. The application
of this approach to more realistic cosmologies (and the relation to
quintessence models \cite{cwquint,quint}) will be discussed elsewhere.
%
%
\section{The general framework} \label{3}
\subsection{The modified Einstein equation}
Our starting point is the modified Einstein-Hilbert action
$S_{\text{\scriptsize mEH}} \left[ g, G, \Lambda \right]$ of eq.\ \eqref{2.1}
which promotes Newton's constant and the cosmological
constant to scalar functions on spacetime.
In this setting $G \left( x \right)$ and $\Lambda \left( x \right)$ are 
arbitrary, externally prescribed functions which are assumed to 
have no functional
dependence on the metric a priori. In fact, the functional derivative of
$S_{\text{\scriptsize mEH}}$ with respect to $g_{\mu \nu}$ is given by\footnote
{Our curvature
conventions are $R^{\sigma}_{~\rho \mu \nu} = - \partial_{\nu}
\Gamma_{\mu \rho}^{~\,\sigma} + \cdots$, $R_{\mu \nu} = R^{\sigma}_{~\mu \sigma \nu}$.
The metric signature is $\left( -+++ \right)$. Frequently we abbreviate
$\left( D G \right)^{2} \equiv 
g^{\mu \nu} \, D_{\mu} G \, D_{\nu} G$ and 
$D^{2} G \equiv 
g^{\mu \nu} \, D_{\mu} D_{\nu} G$.
}
\begin{align}
\frac{2}{\sqrt{-g}} \,
\frac{\delta S_{\text{mEH}}\ \left[ g,G,\Lambda \right]}
{\delta g_{\mu \nu} \left( x \right)} & = 
- \frac{1}{8 \pi G \left( x \right)} \, 
\left( G^{\mu \nu} + g^{\mu \nu} \, \Lambda - 
\Delta t^{\mu \nu} \right),
\end{align}
with $G_{\mu \nu} \equiv R_{\mu \nu} - \frac{1}{2} \, g_{\mu \nu} \, R$ the 
usual Einstein tensor. The $x$-dependence of Newton's constant
gives rise to the tensor
\begin{align}
\begin{split}
\Delta t_{\mu \nu} & \equiv
G \left( x \right) \, \left( D_{\mu} D_{\nu} 
- g_{\mu \nu} \, D^{2} \right) \, \frac{1}{G \left( x \right)} \\
& \equiv \frac{1}{G^{2}}
\Bigl \{ 2 \, D_{\mu} G \, D_{\nu} G
- G \, D_{\mu} D_{\nu} G
- g_{\mu \nu} \, \left[ 2 \, \left( D G \right)^{2}
- G \, D^{2} G \right] \Bigr \}.
\end{split} \label{2.3}
\end{align}
For later use we also note its trace and covariant divergence:
\begin{align}
\Delta t_{\mu}^{~\mu} & =
\frac{3}{G^{2}} \, \left[ G \, D^{2} G 
- 2 \, \left( D G \right)^{2} \right], \label{2.4} \\
D_{\mu} \Delta t^{\mu \nu} & =
\frac{1}{G} \, D_{\mu} G \, \left( \Delta t^{\mu \nu} - R^{\mu \nu} 
\right).  \label{2.5}
\end{align}

We introduce an arbitrary set of matter fields $A \left( x \right)$ minimally
coupled to gravity. Their dynamics is governed by the action
$S_{\text{\scriptsize M}} \left[ g,A \right]$ which gives rise to the
energy-momentum tensor of the matter system in the usual way:
\begin{align}
T^{\mu \nu} = \frac{2}{\sqrt{-g}} \, 
\frac{\delta S_{\text{M}} \left[ g,A \right]}
{\delta g_{\mu \nu} \left( x \right)}.
\end{align}
The action $S_{\text{\scriptsize M}}$ is assumed to be invariant under general
coordinate transformation. As a result, $T_{\mu \nu}$ is conserved when
$A$ satisfies its equation of motion, 
$\delta S_{\text{\scriptsize M}} / \delta A =0$:
\begin{align}
D_{\mu} T^{\mu \nu} =0. \label{2.7}
\end{align}

Furthermore, we allow for an action $S_{\theta} \left[ g,G, \Lambda \right]$
and a corresponding energy-momentum tensor
\begin{align}
\theta^{\mu \nu} = \frac{2}{\sqrt{-g}} \, 
\frac{\delta S_{\theta} \left[ g,G,\Lambda \right]}
{\delta g_{\mu \nu} \left( x \right)} \label{2.8}
\end{align}
which is supposed to describe the 4-momentum carried by the fields
$G \left( x \right)$ and $\Lambda \left( x \right)$. It
vanishes for constant fields therefore. The structure of
$S_{\theta}$ and $\theta_{\mu \nu}$ is not completely fixed by general
principles. In fact, one of the main topics of the present paper is a 
detailed discussion of the mathematical consistency requirements constraining
the form of $\theta_{\mu \nu}$, and of the physical implications of 
various choices for $\theta_{\mu \nu}$. The only general properties
which are assumed throughout are (1), $S_{\theta}$ is independent of
the matter fields, (2), $S_{\theta}$ is invariant under general 
coordinate transformations provided $G$ and $\Lambda$ are transformed
as scalars, and (3), $\theta_{\mu \nu}$ vanishes for $G, \Lambda = const$
(otherwise we would modify the ordinary Einstein equation).

Thus the total action for the system under consideration is
\begin{align}
S_{\text{tot}} & = S_{\text{mEH}} \left[ g,G,\Lambda \right]
+ S_{\text{M}} \left[ g,A \right] + S_{\theta} \left[ g,G,\Lambda \right].
\end{align}
Varying $S_{\text{\scriptsize tot}}$ with respect to the metric we obtain
a modified form of Einstein's equation:
\begin{align}
G_{\mu \nu} & =
- \Lambda \, g_{\mu \nu} + 8 \pi G \, 
\left( T_{\mu \nu} + \Delta T_{\mu \nu} + \theta_{\mu \nu} \right).
\label{2.10}
\intertext{Frequently we shall write it in the form}
G_{\mu \nu} & =
- \Lambda \, g_{\mu \nu} + 8 \pi G \, T_{\mu \nu}
+ \Delta t_{\mu \nu} + \vartheta_{\mu \nu} \label{2.11}
\end{align}
with the convenient definitions
\begin{align}
\vartheta_{\mu \nu} & \equiv 8 \pi G \, \theta_{\mu \nu}, \label{2.12}\\
\Delta t_{\mu \nu} & \equiv 8 \pi G \, \Delta T_{\mu \nu}. \label{2.13}
\end{align}
An equivalent form of the field equation is obtained from \eqref{2.11}
by contracting with $g^{\mu \nu}$:
\begin{align}
R_{\mu \nu} & = Q_{\mu \nu} + \widetilde{\Delta t}_{\mu \nu} 
+ \widetilde \vartheta_{\mu \nu}. \label{2.14}
\end{align}
The source terms on the RHS of \eqref{2.14} are given by
\begin{align}
Q_{\mu \nu} & \equiv \Lambda \, g_{\mu \nu} 
+ 8 \pi G \, \left( T_{\mu \nu} - \tfrac{1}{2} \, g_{\mu \nu} \, T \right), 
\label{2.15}\\
\widetilde{\Delta t}_{\mu \nu} & \equiv \Delta t_{\mu \nu}
- \tfrac{1}{2} \, g_{\mu \nu} \, \Delta t_{\alpha}^{~\alpha}, 
\label{2.16}\\
\widetilde \vartheta_{\mu \nu} & \equiv \vartheta_{\mu \nu}
- \tfrac{1}{2} \, g_{\mu \nu} \, \vartheta_{\alpha}^{~\alpha}.
\label{2.17}
\end{align}
Here and in the following we write $T \equiv T_{\mu}^{~\mu}$ for the trace
of the energy-momentum tensor in the matter sector.

Einstein's equation is coupled to the equations of motion of the matter
system,
\begin{align}
\frac{\delta S_{\text{tot}}}{\delta A} = \frac{\delta S_{\text{M}}}{\delta A}
=0. \label{2.17'}
\end{align}

We stress that \eqref{2.17'} and Einstein's
equation $\delta S_{\text{\scriptsize tot}} / \delta g_{\mu \nu} =0$
are the only field equations to be derived from $S_{\text{\scriptsize tot}}$;
there are no analogous equations for $G$ and $\Lambda$ such as
$\delta S_{\text{\scriptsize tot}} / \delta G =0=\delta S_{\text{\scriptsize tot}} / \delta \Lambda$.
This is a key difference between our approach and standard Brans-Dicke
type theories. As we emphasized already,
 $G \left( x \right)$ and $\Lambda \left( x \right)$
are \textit{externally prescribed} functions in our case
which are determined by the RG equations for $G \left( k \right)$ and
$\Lambda \left( k \right)$ and an appropriate cutoff identification. 
For this reason the status of $G$ and $\Lambda$
is different from that of a scalar matter field $A$.
For instance, their energy-momentum tensor $\theta_{\mu \nu}$
is not conserved even though $S_{\theta}$ is invariant under 
general coordinate transformations. 
To see this, consider an infinitesimal diffeomorphism generated by an
arbitrary vector field $V^{\mu}$. The invariance of $S_{\theta}$ implies
that, to first order in $V^{\mu}$,
\begin{align}
S_{\theta} \left[ 
g_{\mu \nu} + D_{\mu} V_{\nu} + D_{\nu} V_{\mu},~
G + V^{\mu} \, D_{\mu} G,~
\Lambda + V^{\mu} \, D_{\mu} \Lambda \right]
= S_{\theta} \left[ 
g_{\mu \nu}, G, \Lambda \right] \label{2.18}.
\end{align}
Using \eqref{2.8} this condition boils down to
\begin{align}
D^{\mu} \theta_{\mu \nu} =
\frac{1}{\sqrt{-g}} \, \left(
\frac{\delta S_{\theta}}{\delta G} \, D_{\nu} G
+ \frac{\delta S_{\theta}}{\delta \Lambda} \, D_{\nu} \Lambda \right).
\label{2.19}
\end{align}
If $G$ and $\Lambda$ were conventional scalars satisfying equations of motion
$\delta S_{\theta} / \delta G = 0 = \delta S_{\theta} / \delta
\Lambda$ the RHS of \eqref{2.19} would vanish and $\theta_{\mu \nu}$ would be
conserved.
This is the standard argument leading to the conservation law \eqref{2.7}.
In the external field problem at hand the functional derivatives of
$S_{\theta}$ with respect to $G$ and $\Lambda$ have no reason to vanish,
however, and $\theta_{\mu \nu}$ is not conserved in general. (A trivial
exception is the choice $S_{\theta} \equiv 0$ which is also investigated below.)
\subsection{The consistency condition}
The modified Einstein equation \eqref{2.11} is subject to a rather
restrictive integrability condition. As a consequence of Bianchi's identities
the divergence of its LHS vanishes identically, $D^{\mu} G_{\mu \nu}
=0$, and so the divergence of the RHS has to vanish, too:
\begin{align}
D^{\mu} \Delta t_{\mu \nu} 
+D^{\mu} \vartheta_{\mu \nu} 
- D_{\nu} \Lambda
+ 8 \pi \, \left( D_{\mu} G \right) T^{\mu}_{~\nu} =0.
\label{2.20}
\end{align}
Einstein's equation admits solutions only if this equation, henceforth
referred to as the ``consistency condition'', is satisfied. For $T_{\mu \nu}$
and $\vartheta_{\mu \nu}$ fixed, it can be regarded as a constraint on 
possible ``backgrounds'' $\left( G \left( x \right), \Lambda \left( x \right)
\right)$ which admit a consistent dynamics of the metric and the matter fields.
Conversely, we could insist on a specific physically motivated background
$\left( G \left( x \right), \Lambda \left( x \right) \right)$.
In this case eq.\ \eqref{2.20} is a condition on the tensor 
$\vartheta_{\mu \nu}$. From this point of view the consistency condition is
highly welcome since, as we shall see, it can reduce the arbitrariness in the
choice of $\theta_{\mu \nu}$ quite significantly.

By virtue of \eqref{2.5}, the first term of the LHS of \eqref{2.20}
is known explicitly,
\begin{align}
\frac{1}{G} \, D^{\mu} G \, \left( \Delta t_{\mu \nu}
- R_{\mu \nu} \right) 
+D^{\mu} \vartheta_{\mu \nu} 
- D_{\nu} \Lambda
+ 8 \pi \, \left( D_{\mu} G \right) T^{\mu}_{~\nu} =0
\label{2.21}
\end{align}
where \eqref{2.3} should be inserted for $\Delta t_{\mu \nu}$.
For clarity we shall sometimes refer to \eqref{2.20} or \eqref{2.21} as the 
``consistency condition proper'' or, for a reason which will become clear
in a moment, as the ``off-shell consistency condition''.

With $G$, $\Lambda$ and $\vartheta_{\mu \nu}$ fixed, Einstein's equation and
the consistency condition are two independent sets of equations for 
$g_{\mu \nu}$ which have to be solved simultaneously (together with the
matter field equations).
Therefore it is legitimate to insert one of the equations into the other, and
to use the resulting new equation as the independent one. 
   We take advantage of this
freedom by eliminating the Ricci tensor in the consistency condition 
\eqref{2.21} by means of Einstein's equation in the form \eqref{2.14}. The 
latter yields
\begin{align}
R_{\mu \nu} - \Delta t_{\mu \nu} = 
Q_{\mu \nu} + \widetilde \vartheta_{\mu \nu}
- \tfrac{1}{2} \, g_{\mu \nu} \, \Delta t_{\alpha}^{~\alpha} \label{2.22}.
\end{align}
Inserting \eqref{2.22} with \eqref{2.4} into \eqref{2.21} we find
\begin{align}
\begin{split}
& \frac{3}{2} \, \frac{D_{\nu} G}{G^{3}} \,
\left[ G \, D^{2} G - 2 \, \left( DG \right)^{2} \right]
+ D^{\mu} \vartheta_{\mu \nu}
- \frac{D^{\mu} G}{G} \, \widetilde \vartheta_{\mu \nu} \\
& \phantom{{=}} + 4 \pi \, T \, D_{\nu} G
- \frac{1}{G} \, D_{\nu} \left( G \Lambda \right) =0.
\end{split}
\label{2.23}
\end{align}
Very often this alternative form of the consistency condition is more easily
analyzed than the original one. 
Eqs.\ \eqref{2.20} and \eqref{2.23} are equivalent ``on-shell'', i.\,e.\ only
when $g_{\mu \nu}$ satisfies its field equation. We shall therefore refer
to \eqref{2.23} as the ``on-shell consistency condition''.
\subsection{The Brans-Dicke type $\boldsymbol{\theta}$-tensor} \label{2.C}
Before continuing the general discussion let us look at an important special
case. It is characterized by the absence on any matter, $T_{\mu \nu} =0$,
and a background $\left( G \left( x \right), \Lambda \left( x \right) \right)$
with $G \left( x \right) \Lambda \left( x \right) = const$ which is realized
in the fixed point regime, for instance.

In this special case the ``on-shell'' consistency condition \eqref{2.23}
reduces to
\begin{align}
D^{\mu} \vartheta_{\mu \nu}
- \frac{D^{\mu} G}{G} \, \widetilde \vartheta_{\mu \nu}
+ \frac{3}{2} \, \frac{D_{\nu} G}{G^{3}} \,
\left[ G \, D^{2} G - 2 \, \left( DG \right)^{2} \right] =0
\label{2.24}
\end{align}
This is an equation for $\vartheta_{\mu \nu}$ as a function of $G$ and its
derivatives. In order to analyze it, it is convenient to introduce the field
\begin{align}
\psi \left( x \right) \equiv - \ln \left[ G \left( x \right)
/ \, \overline{G} \, \right] \label{2.25}
\end{align}
so that $G = \overline{G} \, e^{- \psi}$ where $\overline{G}$ is an arbitrary
constant reference value. In terms of $\psi$, eq.\ \eqref{2.24} reads
\begin{align}
D^{\mu} \vartheta_{\mu \nu}
+ \widetilde \vartheta_{\mu \nu} \, D^{\mu} \psi
+ \frac{3}{2} \, D_{\nu} \psi\,
\Bigl[ \left( D \psi \right)^{2} + D^{2} \psi \Bigr] =0.
\label{2.26}
\end{align}
In Appendix \ref{appendix B} we show that the \textit{unique}
tensor satisfying \eqref{2.26} identically in $\psi$ and 
vanishing for $\psi = const$ is given by
\begin{align}
\vartheta_{\mu \nu}^{\text{\scriptsize BD}}
& = - \frac{3}{2} \, \Bigl[ D_{\mu} \psi \, D_{\nu} \psi
- \tfrac{1}{2} \, g_{\mu \nu} \, \left( D \psi \right)^{2} \Bigr] 
\label{2.27}\\
& = - \frac{3}{2 \, G^{2}} \,
\Bigl[ D_{\mu} G \, D_{\nu} G
- \tfrac{1}{2} \, g_{\mu \nu} \, \left( D G \right)^{2} \Bigr].
\label{2.28}
\end{align}

This example nicely 
illustrates the power of the consistency condition: it has completely fixed
the form of $\theta_{\mu \nu}^{\text{\scriptsize BD}} 
= \vartheta_{\mu \nu}^{\text{\scriptsize BD}} / 8 \pi G$. However,
this uniqueness property follows only if one demands that $\vartheta_{\mu \nu}$
satisfies \eqref{2.26} identically with respect to $\psi$, i.\,e.\ that the
consistency condition is satisfied for \textit{all} background fields
$\psi \left( x \right)$ or $G \left( x \right)$. Actually this is not necessary
in our approach: it is sufficient that the consistency condition is satisfied
by the specific background supplied by the RG equation. If the class of
functions $\psi$ is restricted to have special properties, additional solutions
can exist. If $\psi$ is assumed to solve 
$\left( D \psi \right)^{2} + D^{2} \psi =0$, say, $\vartheta_{\mu \nu}=0$
is a solution of this kind.

Let us look more closely at the tensor
\begin{align}
\theta_{\mu \nu}^{\text{\scriptsize BD}}
= \left( - \frac{3}{2} \right) \, \frac{1}{8 \pi G^{3}} \,
\Bigl[D_{\mu} G \, D_{\nu} G
- \tfrac{1}{2} \, g_{\mu \nu} \, \left( D G \right)^{2} \Bigr].
\label{2.30}
\end{align}
When reexpressed in terms of $\phi \equiv 1 / G$, 
$\theta_{\mu \nu}^{\text{\scriptsize BD}}$ is seen to equal precisely the 
Brans-Dicke energy-momentum tensor 
${\mathcal T}_{\mu \nu}^{\omega}$ of eq.\ \eqref{1.4}
provided one sets $\omega = - 3/2$ there.

Thus it might seem that, at least for this special case, the theory we have
constructed is equivalent to standard Brans-Dicke theory whose coupled
system of field equations for $\phi$ and $g_{\mu \nu}$ is consistent
only if $\phi$ satisfies the Klein-Gordon equation
\begin{align}
\left( 3 + 2 \omega \right) \, D^{2} \phi = 8 \pi \, T. \label{2.31}
\end{align}
For a generic value of $\omega$ this equation determines $\phi$ which
cannot be treated as an externally prescribed field then. What comes to our
rescue here is that $\omega=-3/2$ amounts to the singular limit of
Brans-Dicke theory where \eqref{2.31} degenerates to the statement $T=0$.
This is no longer an equation of motion of $\phi$ but rather a constraint
on the matter system.
In the case at hand the trace of $T_{\mu \nu}$ vanishes trivially, and
\eqref{2.31} is
satisfied in the form $0=0$ for \textit{any} function $D^{2} 
\phi \left( x \right)$. Therefore, as it should be, it does not
fix the form of $\phi \left( x \right)$ in our case.

Even though we are not doing Brans-Dicke theory here, we shall refer to
$\theta_{\mu \nu}^{\text{\scriptsize BD}}$ as the ``$\theta$-tensor of
Brans-Dicke type'' because it has the same structure as 
${\mathcal T}_{\mu \nu}^{\omega}$.

It is easily checked that, via eq.\ \eqref{2.8}, 
$\theta_{\mu \nu}^{\text{\scriptsize BD}}$ can be obtained from the following
action:
\begin{align}
S_{\theta}^{\text{\scriptsize BD}}
 \left[ g,G \right] & = \frac{3}{32 \pi} \, 
\int \!\! \text{d}^{4}x~
\sqrt{-g}
\, \frac{D_{\mu} G \, D^{\mu} G}{G^{3}} \label{2.32} \\
& = \frac{3}{32 \pi \overline{G}} \,
\int \! \! \text{d}^{4} x ~
\sqrt{-g} \,  e^{\psi} \, 
D_{\mu} \psi \, D^{\mu} \psi. \label{2.33}
\end{align}
Note that $S_{\theta}^{\text{\scriptsize BD}}$ does not depend on $\Lambda$.

As we argued already, $\theta_{\mu \nu}$ is not conserved in general. For the
case of $\theta_{\mu \nu}^{\text{\scriptsize BD}}$ this can be demonstrated
explicitly by taking the divergence of \eqref{2.30}:
\begin{align}
D^{\mu} \theta_{\mu \nu}^{\text{\scriptsize BD}} =
\frac{3}{32 \pi G^{4}} \, \Bigl[ 3 \, \left( D G \right)^{2}
- 2 \, G \, D^{2} G \Bigr] \, D_{\nu} G. \label{2.34}
\end{align}
For an alternative proof of this relation one can insert the action 
\eqref{2.32} into eq.\ \eqref{2.19}. Note that the non-conservation of
$\theta_{\mu \nu}^{\text{\scriptsize BD}}$ is not simply due to the
$x$-dependence of $G$ in \eqref{2.12}; the tensor
$\vartheta_{\mu \nu}^{\text{\scriptsize BD}}$, too, is not conserved:
\begin{align}
D^{\mu} \vartheta_{\mu \nu}^{\text{\scriptsize BD}} =
\frac{3}{2 \, G^{3}} \, \Bigl[ \left( D G \right)^{2}
- G \, D^{2} G \Bigr] \, D_{\nu} G. \label{2.35}
\end{align}

Next we resume the discussion of arbitrary $T_{\mu \nu}$'s and backgrounds.
\subsection{Special classes of solutions} \label{2.D}
Up to this point, our discussion of RG improved action functionals has lead
to a coupled system of effective field equations consisting of
(1) Einstein's equation \eqref{2.11},
(2) the equation of motion of the matter fields \eqref{2.17'}, and
(3) the consistency condition \eqref{2.21}.

Clearly it is very difficult in general to find solutions to this coupled
system. Therefore, to start with, we discuss various classes of solutions 
which result from making certain simplifying assumptions. The specification
of a class involves (i) a choice of $\theta_{\mu \nu}$, (ii) assumptions
about the external fields $G \left( x \right)$, $\Lambda \left( x \right)$,
and (iii) assumptions about the matter system. Some of the classes have
quite remarkable properties which will be explored further in 
Subsection \ref{2.E}. In Section \ref{4} on cosmology we impose the 
symmetry condition of homogeneity and isotropy which allows us to find
explicit examples for all classes.

In this paper we investigate two choices of the $\theta$-tensor:
$\theta_{\mu \nu} \equiv 0$ and $\theta_{\mu \nu} = 
\theta_{\mu \nu}^{\text{\scriptsize BD}}$. The motivation for using the
Brans-Dicke energy-momentum tensor is twofold: It 
enjoys the uniqueness property discussed in Subsection \ref{2.4},
and it allows for a comparison of standard Brans-Dicke theory
with our ``external field Brans-Dicke theory''.

To be specific, we shall deal with the following special cases:
\begin{description}
\item [Class I:] This class is defined by the choice $\theta_{\mu \nu} \equiv
0$. Solutions of Class I satisfy
\begin{subequations} \label{2.36}
\begin{gather}
G_{\mu \nu} = - \Lambda \, g_{\mu \nu} + \Delta t_{\mu \nu} 
+ 8 \pi G \, T_{\mu \nu} \label{2.36a}\\
\frac{3}{2} \, \frac{D_{\nu} G}{G^{2}} \, 
\left[ G \, D^{2} G - 2 \, \left( D G \right)^{2} \right]
+ 4 \pi G \, T \, D_{\nu} G - D_{\nu} \left( G \Lambda \right)
=0 \label{2.36b}
\end{gather}
\end{subequations}
along with $\delta S_{\text{M}} / \delta A =0$.
\item [Class II:] This class is defined by an identically vanishing 
cosmological constant and a matter system whose energy-momentum tensor
has a vanishing trace $T \equiv T_{\mu}^{~\mu}$, at least ``on-shell'':
\begin{subequations} \label{2.37}
\begin{align}
\Lambda =0 \quad \text{and} \quad T=0. \label{2.37a}
\end{align}
As a result, in Class II the on-shell 
consistency condition \eqref{2.23} reduces to 
eq.\ \eqref{2.24}. In Subsection \ref{2.C} we saw that the unique tensor 
solving this latter equation identically in $\psi$ is
$\vartheta_{\mu \nu}^{\text{\scriptsize BD}}$. Allowing for special properties
of $\psi$ 
there exist more general solutions, but as one of the defining properties
of this specific class we include
\begin{align}
\theta_{\mu \nu} = \theta_{\mu \nu}^{\text{\scriptsize BD}} \label{2.37b}
\end{align}
\end{subequations}
into the definition of Class II. Thanks to \eqref{2.37a} and \eqref{2.37b} the
on-shell consistency condition is satisfied by construction, and it remains to
solve Einstein's equation together with the matter field equation of motion.
\item [Class III:] This class is characterized by
\begin{subequations} \label{2.38}
\begin{gather}
\theta_{\mu \nu} = \theta_{\mu \nu}^{\text{\scriptsize BD}}
\quad \text{and} \quad
\Lambda \neq 0. \label{2.38a}
\end{gather}
The on-shell consistency condition \eqref{2.23} is not satisfied 
automatically, but it simplifies considerably. As a result of the choice
$\theta_{\mu \nu} = \theta_{\mu \nu}^{\text{\scriptsize BD}}$, its first
line vanishes identically and it remains to impose
\begin{gather}
4 \pi G \, T \, \partial_{\mu} G = \partial_{\mu} \left( G \Lambda \right)
\label{2.38b}.
\end{gather}
\end{subequations}
\end{description}

It is convenient to distinguish two sub-classes of the Class III:
\begin{description}
\item [Class IIIa:] This sub-class corresponds to the special case when
\begin{gather}
T=0 \quad \text{and} \quad G \Lambda = const. \label{2.39}
\end{gather}
In this case the residual consistency condition \eqref{2.38b} is solved
automatically, its LHS and RHS both being zero.
\item [Class IIIb:] In this sub-class the residual consistency condition is 
satisfied in a non-trivial way, i.\,e.\ not in the form ``$0=0$'':
\begin{gather}
4 \pi G \, T \, \partial_{\mu} G = \partial_{\mu} \left( G \Lambda \right)
\neq 0. \label{2.40}
\end{gather}
\end{description}

It is quite intriguing that the RHS of the residual consistency condition
\eqref{2.38b} vanishes precisely when the product $G \Lambda$ is constant, as
it is the case in the fixed point regime, for instance. Assuming 
$\partial_{\mu} G \neq 0$, the LHS vanishes only if $T_{\mu \nu}$ is 
traceless, i.\,e.\ when the matter system is described by a quantum conformal
field theory with vanishing trace anomaly, for instance. What eq.\ \eqref{2.38b}
tells us is that when gravity is at a critical point (fixed point regime) so 
must be the matter fields.
In this situation the combined gravity plus matter system can be regarded as
 a kind of scale invariant ``critical phenomenon''. At least when we use
$\theta_{\mu \nu}^{\text{\scriptsize BD}}$, only traceless matter can be
coupled to gravity at its UV fixed point.

Henceforth we shall assume that the traceless $T_{\mu \nu}$ of the Classes
II and IIIa originates from a Weyl-invariant matter action so that,
for any function $\sigma \left( x \right)$,
\begin{align}
S_{\text{M}} \left[ e^{2 \sigma} g_{\mu \nu}, ~
e^{-2 \Delta_{A} \sigma}
A \right] = S_{\text{M}} \left[ g_{\mu \nu}, A \right]. \label{2.41} 
\end{align}
(For simplicity we consider only a single matter field $A$ with Weyl weight
$\Delta_{A}$.) The infinitesimal form of \eqref{2.41} reads
\begin{align}
T_{\mu}^{~\mu} = \frac{2 \Delta_{A}}{\sqrt{-g}} \,
\frac{\delta S_{\text{M}}}{\delta A} \, A
\label{2.42}
\end{align}
implying that $T_{\mu \nu}$ is indeed traceless when $A$ is on-shell.
\subsection{Solutions from a Weyl transformation} \label{2.E}
In this Subsection we discuss a very powerful tool for analyzing the
Classes II and IIIa. As we shall see, solutions of these types can be
obtained by simply Weyl-transforming solutions of the standard Einstein
equations with constant $G$ and $\Lambda$.

There exists an extensive literature on the use of Weyl transformations
in gravitational theories and on the physical interpretation of the conformal
frames they connect \cite{faraoui}. In the present paper, the Weyl
transformations and the metrics $\gamma_{\mu \nu}$ they lead to (see below)
should be regarded merely a technical tool for generating solutions of the
modified field equations. They have no direct physical significance.
It should also be mentioned that the standard discussion of Weyl rescalings
applied to Brans-Dicke theory \cite{faraoui,maeda} does not apply in our
case since it breaks down at the singular point $\omega = -3/2$ we are
working at.

We start by picking two fixed reference values 
$\overline{\Lambda}$ and $\overline{G}$ of the cosmological and Newton's
constant, respectively, and we introduce the conventional Einstein-Hilbert
actions
\begin{align}
\overline{S}_{\text{EH}} \left[ g  \right] & \equiv
- \frac{1}{16 \pi \overline{G}} \int \!\! \text{d}^{4} x~
\sqrt{-g} \, \Bigl[ - R \left( g \right) + 2 \, \overline{\Lambda} \, \Bigr], 
\label{2.43}\\
\overline{S}_{\text{EH}}^{\,0} \left[ g  \right] & \equiv
\frac{1}{16 \pi \overline{G}} \int \!\! \text{d}^{4} x~
\sqrt{-g} \, R \left( g \right).
\label{2.44}
\end{align}
For the corresponding functionals with the matter action included we write
\begin{align}
\overline{S}_{\text{tot}} \left[ g,A  \right] & \equiv
\overline{S}_{\text{EH}} \left[ g  \right] + 
S_{\text{M}} \left[ g,A  \right], 
\label{2.45}\\
\overline{S}_{\text{tot}}^{\, 0} \left[ g,A  \right] & \equiv
\overline{S}_{\text{EH}}^{\, 0} \left[ g  \right]
+ S_{\text{M}} \left[ g,A  \right]. 
\label{2.46}
\end{align}
We assume that $S_{\text{M}}$ is Weyl invariant.

Let us focus on \textbf{Class II} first. Here, by definition, $\Lambda =0$
so that the total action reads
\begin{align}
S_{\text{tot}} \left[ g, G, A  \right] & = 
S_{\text{mEH}} \left[ g, G, 0 \right]
+ S_{\theta}^{\text{\scriptsize BD}} \left[ g, G \right]
+ S_{\text{M}} \left[ g,A  \right] \label{2.47}
\end{align}
Substituting $1 / G = e^{\psi} / \overline{G}$ in eq.\ \eqref{2.1}
and using \eqref{2.33} we find that
\begin{align}
S_{\text{mEH}} \left[ g, G, 0 \right]
+ S_{\theta}^{\text{\scriptsize BD}} \left[ g, G \right] =
 \frac {1}{16 \pi \overline{G}} \,\int \!\! \text{d}^{4}x~ 
\sqrt{-g}\, e^{\psi} \, 
\Bigl[  R + \tfrac{3}{2} \, D_{\mu} \psi \, D^{\mu} \psi \Bigr].
\label{2.48}
\end{align}
Remarkably, the coefficient of the $\left( D \psi \right)^{2}$-term
in \eqref{2.48} is precisely such that this term can be absorbed into the
$R$-term by means of a Weyl rescaling of the metric\footnote{
Recall that the transformation $g_{\mu \nu}^{\prime}=e^{2\sigma} \,
g_{\mu \nu}$ gives rise to
$\int \!\! \text{d}^{4}x~ \sqrt{-g^{\prime}} = 
\int \!\! \text{d}^{4}x~ \sqrt{-g} \, e^{4 \sigma}$ and
$\int \!\! \text{d}^{4}x~ \sqrt{-g^{\prime}} \, R \left( g^{\prime} \right) = 
\int \!\! \text{d}^{4}x~ \sqrt{-g} \, e^{2 \sigma} \, 
\Bigl[ R \left( g \right) + 6 \, D_{\mu} \sigma \, D^{\mu}
\sigma \Bigr]$.
In obtaining the second relation an integration by parts has been performed
and the surface term has been dropped.
} with $\sigma = \psi / 2$:
\begin{align}
S_{\text{mEH}} \left[ g, G, 0 \right]
+ S_{\theta}^{\text{\scriptsize BD}} \left[ g, G \right] 
= \overline{S}_{\text{EH}}^{\,0} \left[ \frac{\overline{G}}{G} \,
g_{\mu \nu} \right]
\equiv \overline{S}_{\text{EH}}^{\,0} \left[ \gamma_{\mu \nu} \right].
\label{2.49}
\end{align}
The rescaled metric is $\gamma_{\mu \nu} = e^{\psi} \, g_{\mu \nu}$, or
\begin{subequations} \label{2.50}
\begin{align}
\gamma_{\mu \nu} \left( x \right) = \frac{\overline{G}}{G \left( x \right)}
\, g_{\mu \nu} \left( x \right). \label{2.50a}
\end{align}
Thus the sum of $S_{\text{mEH}}$ and $S_{\theta}^{\text{\scriptsize BD}}$
equals the \textit{standard} Einstein-Hilbert action with a constant value
of Newton's constant. The only place where the RG improvement manifests
itself is the conformal factor $\overline{G} / G \left( x \right)$
which gets attached to the metric. The possibility of performing this
Weyl transformation is directly related to the fact that we work in the 
singular limit of Brans-Dicke theory where $\omega = - 3/2$ \cite{blago}.
In standard Brans-Dicke theory there is no such possibility.

The matter action is Weyl invariant for any $\sigma$, and so in particular
for $\sigma = \psi / 2$. Hence it follows that 
$S_{\text{M}} \left[ g_{\mu \nu}, A \right]=
S_{\text{M}} \left[ \gamma_{\mu \nu}, {\mathcal A} \right]$
with the rescaled matter field
\begin{align}
{\mathcal A} \left( x \right) = \left[ \frac{\overline{G}}{G \left( x \right)}
\right]^{- \Delta_{A}} \, A \left( x \right). \label{2.50b}
\end{align}
\end{subequations}
As a consequence, the total action \eqref{2.47} for a position-dependent
$G$ reduces to \eqref{2.46} for a constant $G$, up to a rescaling of the
fields according to \eqref{2.50}:
\begin{align}
S_{\text{tot}} \left[ g, G, A  \right] = 
\overline{S}_{\text{tot}}^{\, 0} \left[ \gamma, {\mathcal A}  \right].
\label{2.51}
\end{align}

If a configuration $\left( g_{\mu \nu}, A \right)$ 
is a stationary point of the functional
 $S_{\text{tot}} \left[ g, G, A  \right]$, the related 
configuration $\left( \gamma_{\mu \nu}, {\mathcal A} \right)$
is a stationary point of
$\overline{S}_{\text{tot}}^{\, 0} \left[ \gamma, {\mathcal A}  \right]$,
i.\,e.\ it is a solution of the standard constant-$G$, 
$\Lambda =0$-Einstein equation
\begin{subequations} \label{2.52}
\begin{gather}
G_{\mu \nu} \left( \gamma \right) = 
8 \pi \overline{G} \, \, T_{\mu \nu} \left( {\mathcal A}, \gamma \right)
\label{2.52a}
\intertext{and the matter equation}
\delta \overline{S}_{\text{tot}}^{\, 0} / \delta {\mathcal A} =0.
\label{2.52b}
\end{gather}
\end{subequations}
This observation provides us with a very efficient technique for obtaining the
solutions of Class II: we take any solution
$\left( \gamma_{\mu \nu}, {\mathcal A} \right)$
of the much simpler system \eqref{2.52}{ and invert the Weyl transformation
\eqref{2.50} in order to find $\left( g_{\mu \nu}, A \right)$:
\begin{subequations} \label{2.53}
\begin{align}
g_{\mu \nu} \left( x \right) & = 
\frac{G \left( x \right)}{\overline{G}} \, \gamma_{\mu \nu} \left( x \right)
\label{2.53a} \\
A \left( x \right) & =
\left[ \frac{ G \left( x \right)}{\overline{G}} \right]^{- \Delta_{A}}
\, {\mathcal A} \left( x \right). \label{2.53b}
\end{align}
\end{subequations}
For $G \left( x \right)$ regular, the Weyl transformation is invertible and
\textit{all} solutions can be found in this manner.

Until now we assumed that the cosmological constant vanishes. If we allow
for an arbitrary function $\Lambda \left( x \right)$, the $\Lambda$-term
in $S_{\text{mEH}}$,
\begin{align*}
\frac{1}{16 \pi} \, \int \! \! \text{d}^{4} x ~
\sqrt{-g} \, \biggl \{ -2 \, 
\frac{\Lambda \left( x \right)}{G \left( x \right)} \biggr \},
\end{align*}
makes it impossible to use a 
Weyl transformation in order to convert $G \left( x 
\right)$ to $\overline{G}$ everywhere. However, there is one exception
to this rule, namely when $\Lambda \left( x \right)$ is proportional to 
$1 / G \left( x \right)$, i.\,e.\ when the product
$G \left( x \right) \Lambda \left( x \right)$ is constant. This situation
corresponds precisely to the solutions of \textbf{Class IIIa} to which
we turn next. We write the constant product in the form
\begin{align}
G \left( x \right) \Lambda \left( x \right) = 
\overline{G} \, \overline{\Lambda} \label{2.54},
\end{align}
whence $\Lambda \left( x \right) / G \left( x \right)=
e^{2 \psi} \, \overline{\Lambda} / \, \overline{G}$.
As a result, the $\Lambda$-term, too, can be brought to its constant-$G$,
constant-$\Lambda$ form by the Weyl rescaling \eqref{2.50a}:
\begin{align}
\frac{1}{16 \pi} \, \int \! \! \text{d}^{4} x ~
\sqrt{-g} \, \biggl \{ -2 \, 
\frac{\Lambda \left( x \right)}{G \left( x \right)} \biggr \} =
\frac{1}{16 \pi \overline{G}} \, \int \! \! \text{d}^{4} x ~
\sqrt{- \gamma} \Bigl \{ -2 \, \overline{\Lambda} \Bigr \}.
\label{2.55}
\end{align}
Again we observe that the case $G \Lambda = const$ which is special from
the RG point of view because it is realized in the fixed point regime enjoys 
very special properties also with respect to the gravitational actions and
field equations. Combining \eqref{2.51} with \eqref{2.55} we see that for the
Class IIIa the total action can be reduced to the one with the $x$-independent
$G$ and $\Lambda$:
\begin{align}
S_{\text{tot}} \left[ g, G, \Lambda, A  \right] = 
\overline{S}_{\text{tot}} \left[ \gamma, {\mathcal A}  \right].
\label{2.56}
\end{align}
The stationary points $\left( g_{\mu \nu}, A \right)$ of
$S_{\text{tot}} \left[ g, G, \Lambda, A  \right]$ are related to the 
stationary points  
$\left( \gamma_{\mu \nu}, {\mathcal A} \right)$
of
$\overline{S}_{\text{tot}} \left[ \gamma, {\mathcal A}  \right]$
by the same transformations as above, eqs.\ \eqref{2.50}.
The latter are solutions to the conventional constant-$G$, constant-$\Lambda$
field equations
\begin{subequations} \label{2.57}
\begin{gather}
G_{\mu \nu} \left( \gamma \right) = 
- \overline{\Lambda} \, \gamma_{\mu \nu} + 
8 \pi \overline{G} \, \, T_{\mu \nu} \left( {\mathcal A}, \gamma \right)
\label{2.57a}
\\
\delta \overline{S}_{\text{tot}} / \delta {\mathcal A} =0.
\label{2.57b}
\end{gather}
\end{subequations}

Thus we conclude that the solutions of the Class IIIa can be obtained by
Weyl-transforming the solutions of the system \eqref{2.57} according to
eqs.\ \eqref{2.53}. There is no corresponding simplification in Class IIIb.

Clearly the Weyl transformation $\gamma_{\mu \nu} = e^{\psi} \, g_{\mu \nu}$
can always be performed, whatever is $\theta_{\mu \nu}$, $T_{\mu \nu}$,
$G$, and $\Lambda$. It takes us from the ``$g_{\mu \nu}$-frame'' to the
``$\gamma_{\mu \nu}$-frame'' in which $\gamma_{\mu \nu}$ is the independent
variable and where the modified Einstein equation reads
\begin{align}
\begin{split}
G_{\mu \nu} \left( \gamma \right) & =
\tfrac{3}{2} \, \left( 1 - \varepsilon \right) \,
\Bigl[ D_{\mu} \psi \, D_{\nu} \psi 
- \tfrac{1}{2} \, \gamma_{\mu \nu} \gamma^{\alpha \beta} \,
D_{\alpha} \psi \, D_{\beta} \psi \Bigr]
\\
& \phantom{{==}}
+ \Bigl[ - \Lambda \left( x \right) \, \gamma_{\mu \nu}
+ 8 \pi \overline{G} \, T_{\mu \nu} \left( A, e^{-\psi} \gamma \right)
\Bigr] \, e^{-\psi}
\end{split}
\label{2.100}
\end{align}
In writing down \eqref{2.100} we adopted the choice
$\theta_{\mu \nu} = \varepsilon \, \theta_{\mu \nu}^{\text{\scriptsize BD}}$
where $\varepsilon$ is an arbitrary real constant, with $\varepsilon=0$
and $\varepsilon=1$ being particularly interesting cases, of course. Only in 
the classes II and IIIa $\psi$ drops out from the RHS of \eqref{2.100}.
In general this Einstein equation is reminiscent of ordinary gravity plus
a massless Klein-Gordon field. While occasionally this analogy is helpful
for generating solutions, it is quite deceptive from the physical point of
view because physical lengths are still measured by the metric $g_{\mu \nu}$
and not by $\gamma_{\mu \nu}$. For instance, when one applies this formalism
to black holes \cite{holger2} one would like the distance function 
$d \left( r \right)$ appearing in the cutoff identification \eqref{1.11}
to be computed from the actual metric of the improved spacetime, i.\,e.\ from
$g_{\mu \nu}$. If $g_{\mu \nu}$ is to be regarded as the product 
$e^{\psi} \, \gamma_{\mu \nu}$ the situation becomes very involved because
then the cutoff identification we insert into $G \left( k \right)$, besides
$\gamma_{\mu \nu}$, becomes explicitly dependent on $\psi \left( x \right)$,
i.\,e.\ $G \left( x \right)$, itself. In order to avoid complications of this
kind we shall mostly employ the physical ``$g_{\mu \nu}$-frame''.

In Brans-Dicke jargon the $g_{\mu \nu}$- and the 
$\gamma_{\mu \nu}$-frame are called the Jordan and the Einstein frame, 
respectively. There is a longstanding debate in the literature about the issue
of which conformal frame is the physical one. Ref.\ \cite{faraoui}
contains a detailed discussion of this problem, and it is argued there
that only the Einstein frame can be physical. A key role in establishing this
argument is played by the positivity of the energy and by the existence and
the stability of a ground state in the Einstein frame. If this discussion 
applied also to the theory of RG improved actions it would give preference
to $\gamma_{\mu \nu}$, rather than $g_{\mu \nu}$, as the physical metric.
However, we emphasize that none of the arguments put forward in 
\cite{faraoui} is applicable to the ``external field Brans-Dicke theory''.
In particular we stress that there can be no doubt that the ``Jordan'' metric
$g_{\mu \nu}$ is the physical one in our case. This is an immediate
consequence of the effective field theory approach we are employing:
$g_{\mu \nu}$, and not $\gamma_{\mu \nu}$, is the average of the microscopic
metric, the variable of integration in the path-integral over all metrics
\cite{mr}.
By including matter terms into the flow equation it is obvious that it is
$g_{\mu \nu}$ which enters the conservation law $D_{\mu} T^{\mu \nu} =0$
and determines the geodesics of test particles. The standard discussion
does not apply here for a variety of reasons: 
(i) As we mentioned already, $\omega =-3/2$ is a highly singular point.
(ii) In the standard case, the scalar field $G (x)$ exists in an unambiguous,
i.\,e.\ ``process independent'' way. Not so in our case: A priori $G$ is
neither position nor time but rather \textit{scale} dependent, and only under
very special conditions, and not everywhere in spacetime, the $k$-dependence
can be translated to a $x$-dependence. This invalidates, for instance,
standard discussions of the equivalence principle and the ``universality'' of
the gravitational interaction.\footnote{According to a conservative
interpretation of ``universality'' the present theory does imply 
``nonuniversal'' effects, in the sense that the coupling strength $G (k)$
depends on features such as the size of the bodies involved in the process
considered, or on their relative momentum, say (cf.\ the discussion in
the Introduction). However, in accordance with the terminology in particle
physics, we do not consider it appropriate to call a certain interaction
``nonuniversal'' just because the corresponding coupling displays
RG running.}
(iii) The positivity of the action and the properties of the ground state
must be checked on the basis of the exact 
$\Gamma \equiv \Gamma_{k=0}$ \cite{avactrev}. Since only an approximate form of 
$\Gamma_{k}$, $k >0$ is known, we are clearly not yet in a position to address
questions about the vacuum of Quantum Einstein Gravity.
%
%
\section{Cosmological evolution equations} \label{4}
In the remaining sections of this paper we apply the approach developed
in Section \ref{3} to the cosmology of homogeneous and isotropic Universes.

Because of this symmetry requirement, the metric can be brought to the
Robertson-Walker form
\begin{gather}
\text{d} s^{2} = - \text{d} t^{2} 
+ a^{2} \left( t \right) \, \text{d} \Omega_{K}^{2} \label{4.1}
\intertext{where}
\text{d} \Omega_{K}^{2} \equiv \frac{\text{d} r^{2}}{1-Kr^{2}}
+ r^{2} \, \left( \text{d} \theta^{2} 
+ \sin^{2} \theta \, \text{d} \varphi^{2} \right) \label{4.2}
\end{gather}
is the line element of a maximally symmetric 3-space of positive ($K=+1$),
negative ($K=-1$), or vanishing ($K=0$) curvature, respectively.

The scalar functions $G \left( x \right)$ and $\Lambda \left( x \right)$
are assumed to respect all symmetries. This implies that in  the
$\left( t, r, \theta, \varphi \right)$ coordinate system which we shall use 
throughout they may depend on the cosmological time only: 
$G \left( x \right) = G \left( t \right)$ and
$\Lambda \left( x \right) = \Lambda \left( t \right)$.

In the same coordinate system, the energy-momentum tensor is of the form
\begin{align}
T_{\mu}^{~\nu} = \text{diag} \left[ - \rho, p, p, p \right] \label{4.3}
\end{align}
which is consistent with the symmetries provided the density $\rho$ and
pressure $p$ depend on $t$ only. Here, rather than in terms of a field $A$,
we describe the matter sector in a hydrodynamical language in terms
of a perfect fluid. Because of the strong symmetry constraint, the only 
information which is needed about the matter system is its equation of state
$p = p \left( \rho \right)$.
For the time being we leave it unspecified. As always, $T_{\mu}^{~\nu}$ is 
assumed to be conserved, $D_{\nu} T_{\mu}^{~\nu} =0$, which implies
\begin{gather}
\dot \rho + 3 \, H \, \left( \rho + p \right) =0, \label{4.5}
\end{gather}
where $H \left( t \right) \equiv \dot a \left( t \right) / a \left( t \right)$
denotes the Hubble parameter.

Upon inserting the Robertson-Walker metric into \eqref{2.3} we obtain
\begin{align}
\Delta T_{\mu}^{~\nu} \equiv \frac{1}{8 \pi G} \Delta t_{\mu}^{~\nu}
= \text{diag} \left[ - \Delta \rho, \Delta p, \Delta p, \Delta p \right]
\label{4.6}
\end{align}
where
\begin{subequations} \label{4.7}
\begin{gather}
\Delta \rho = \frac{- \Delta t_{0}^{~0}}{8 \pi G} 
= \frac{3}{8 \pi G} \, \left( \frac{ \dot G}{G} \right) \, H
\label{4.7a}
\intertext{and}
\Delta p = \frac{\Delta t_{i}^{~i}}{8 \pi G} =
\frac{1}{8 \pi G} \, 
\left[ 2 \left( \frac{\dot G}{G} \right)^{2}
- 2 \, \left( \frac{\dot G}{G} \right) \, H 
- \frac{\ddot G}{G} \right] \label{4.7b}
\end{gather}
\end{subequations}
is the density and pressure, respectively, which is contributed by the extra
terms in the Einstein equation due to the $t$-dependence of $G$.
(Eq.\ \eqref{4.7b} holds for any $i=1,2,3$; this index is not summed over.) 
Likewise, every choice of $\theta_{\mu \nu}$ leads to only two independent
functions, $\rho_{\theta}$ and $p_{\theta}$, in a Robertson-Walker
background:
\begin{align}
\theta_{\mu}^{~\nu} = 
\text{diag} \left[ -\rho_{\theta}, p_{\theta}, p_{\theta}, p_{\theta} \right].
\label{4.8}
\end{align}

Plugging the metric \eqref{4.1} into Einstein's equation \eqref{2.10} we 
obtain two independent equations\footnote{With our conventions and notations
the non-zero components of the Einstein tensor are
\begin{align*}
G_{0}^{~0} = - 3 \, \left[ H^{2} + \frac{K}{a^{2}} \right]
\quad \text{and} \quad
G_{i}^{~j} = - \left[ H^{2} + 2 \left( \frac{\ddot a}{a} \right)
+ \frac{K}{a^{2}} \right] \delta_{i}^{~j}.
\end{align*}}: its $00$-component
\begin{subequations} \label{4.9}
\begin{gather}
H^{2} + \frac{K}{a^{2}} = 
\frac{8 \pi}{3} \, G \, \left( \rho + \rho_{\Lambda} + \Delta \rho + 
\rho_{\theta} \right). \label{4.9a}
\intertext{and the $ii$-components which are identical for all
values of the spatial index $i=1,2,3$:}
H^{2} + 2 \, \left( \frac{\ddot a}{a} \right) + \frac{K}{a^{2}} =
- 8 \pi G \, \left( p + p_{\Lambda} + \Delta p + p_{\theta} \right).
\label{4.9b}
\end{gather}
\end{subequations}
In writing down eqs.\ \eqref{4.9} we also introduced the vacuum
energy density and pressure, respectively, which are due to the 
cosmological constant:
\begin{gather}
\rho_{\Lambda} \equiv \frac{\Lambda \left( t \right)}{8 \pi G \left( t 
\right)},
\quad
p_{\Lambda} \equiv - \frac{\Lambda \left( t \right)}{8 \pi G \left( t 
\right)}. \label{4.10}
\end{gather}

We shall find it convenient to define a ``critical'' energy density in the
same way as in standard cosmology,
\begin{align}
\rho_{\text{crit}} \equiv 
\frac{3}{8 \pi G \left( t \right)} \, H^{2} \left( t \right), \label{4.11}
\end{align}
and to refer the matter and vacuum density to $\rho_{\text{crit}}$:
\begin{subequations} \label{4.12}
\begin{gather}
\Omega_{\text{M}} \equiv 
\frac{\rho}{\rho_{\text{crit}}},
 \quad
\Omega_{\Lambda} \equiv 
\frac{\rho_{\Lambda}}{\rho_{\text{crit}}}. \label{4.12a}
\intertext{The analogous relative energy densities due to
$\Delta T_{\mu \nu}$ and
$\theta_{\mu \nu}$ are}
\Delta \Omega \equiv
\frac{\Delta \rho}{\rho_{\text{crit}}}, \quad
\Omega_{\theta} \equiv 
\frac{\rho_{\theta}}{\rho_{\text{crit}}}. \label{4.12b}
\end{gather}
\end{subequations}
In the language of the $\Omega$'s, the modified Friedmann equation, i.\,e.\
the $00$-component \eqref{4.9a}, reads
\begin{align}
K & = \dot a^{2} \left[ \Omega_{\text{M}} + \Omega_{\Lambda}
+ \Delta \Omega + \Omega_{\theta} -1 \right]. \label{4.13}
\end{align}
We observe that in order to obtain a spatially flat, expanding Universe
all four densities, $\Omega_{\text{M}}$, $\Omega_{\Lambda}$, 
$\Delta \Omega$, and $\Omega_{\theta}$, must add up to unity.

As for the consistency condition, its original off-shell form \eqref{2.20}
yields:
\begin{align}
\dot G \,
\left( \rho + \rho_{\Lambda} + \Delta \rho + \rho_{\theta} \right)
+ G \, \frac{\text{d}}{\text{d} t} \left( \rho_{\Lambda} + \Delta \rho 
+ \rho_{\theta} \right) 
+ 3 \, G\, H\, \left[ \Delta \rho
+ \Delta p + \rho_{\theta} + p_{\theta} \right]
= 0.
\label{4.14}
\end{align}

At this point the system of coupled cosmological evolution equations consists
of the $00$-component of Einstein's equation \eqref{4.9a}, its 
$ii$-component \eqref{4.9b}, the off-shell 
consistency condition \eqref{4.14}, and the
equation of state $p=p \left( \rho \right)$. As in standard cosmology, it is 
possible to replace the $ii$-component with the continuity equation
$D_{\mu} T^{\mu \nu} = 0$, i.\,e.\ with \eqref{4.5}, as an 
independent relation.
In Appendix \ref{C} we prove that every cosmology with $\dot a \neq 0$ which satisfies
the $00$-component, the off-shell consistency condition,
 and the continuity equation
automatically also satisfies the $ii$-component of Einstein's
equation. This statement holds true for any choice of the $\theta$-tensor.

As we are now going to discard the $ii$-component as an independent equation
of motion it is no longer guaranteed that the consistency condition proper,
the ``off-shell'' condition \eqref{2.20}, is fully equivalent to the 
``on-shell'' condition \eqref{2.23}. We shall therefore employ the off-shell 
condition in the following. (The on-shell condition is nevertheless useful as
any solution has to satisfy it, of course.)

Let us fix some ``background'' by picking two
functions $G \left( t \right)$ and $\Lambda \left( t \right)$. We would
then like to determine $a \left( t\right)$, $\rho \left( t \right)$, and 
$p \left( t \right)$ from the coupled system of cosmological evolution 
equations consisting of
\begin{align}
\begin{split}
1. &~\text{ the $00$-component of Einstein's equation} \\
2. &~\text{ the off-shell consistency condition} \\
3. &~\text{ the continuity equation} \\
4. &~\text{ the equation of state}
\end{split}
\label{4.15}
\end{align}
Obviously we are trying to determine \textit{three} functions from
\textit{four} independent equations. As a result, an arbitrarily chosen
background $\left( G, \Lambda \right)$ and tensor $\theta_{\mu \nu}$,
generically, will not allow for any solution of the system \eqref{4.15}.
Being over-determined, the system \eqref{4.15}
puts restrictions on $\left( G, \Lambda \right)$ and the $\theta$-tensor.
As we mentioned already in Section \ref{3}, 
this is very welcome because $\theta_{\mu \nu}$ is not completely 
determined by general principles. In the ideal case when
the RG equations yield
a certain trajectory 
$\bigl( G \left( k \right), \Lambda \left( k \right) \bigr)$
and a physically plausible cutoff identification $k = k \left( t \right)$
turns it into a background
$\bigl( G \left( t \right), \Lambda \left( t \right) \bigr)$
for which \eqref{4.15} is indeed soluble, there exists a rather non-trivial
conspiracy of the RG- and field-equations. Being comparatively rare
one is inclined to
ascribe a particularly high 
degree of physical relevance to such ``precious'' solutions.

Next we discuss the special choices $\theta_{\mu \nu}=0$ and
$\theta_{\mu \nu} = \theta_{\mu \nu}^{\text{\scriptsize BD}}$ in turn.
\subsubsection*{(a) The choice $\boldsymbol{\theta_{\mu \nu}=0}$}
For a vanishing $\theta$-tensor the $00$- and $ii$-component of
Einstein's equation read, respectively,
\begin{subequations} \label{4.16}
\begin{gather}
H^{2} + \frac{K}{a^{2}} = 
\frac{1}{3} \, \Lambda + \frac{8 \pi}{3} G \, \rho
+ \left( \frac{\dot G}{G} \right) \, H
\label{4.16a} \\
H^{2} + 2 \, \left( \frac{\ddot a}{a} \right) + \frac{K}{a^{2}}
=
\Lambda - 8 \pi G \, p - 2 \, 
\left( \frac{\dot G}{G} \right)^{2}
+ \frac{\ddot G}{G}+ 2 \, \left( \frac{\dot G}{G} \right) \, H
\label{4.16b}
\end{gather}
\end{subequations}
The off-shell consistency condition is obtained by starting from \eqref{2.21} 
and inserting $\Delta t_{0}^{~0}$ from \eqref{4.7a}, $\theta_{\mu \nu}=0$,
and $R_{0}^{~0}= 3 \, \ddot a / a$ which is true for any 
Robertson-Walker metric. The only non-trivial condition follows from
$\nu=0$:
\begin{gather}
 \dot \Lambda
+ 8 \pi \, \dot G \, \rho
+ 3 \, \left( \frac{\dot G}{G} \right)^{2} \, H
+ 3 \, \left( \frac{ \dot G}{G} \right) \, \left( \frac{\ddot a}{a} \right)= 0.
\label{4.17}
\end{gather}
As a check, it can be verified explicitly that \eqref{4.16a}, \eqref{4.17},
and the continuity equation imply \eqref{4.16b}.
\subsubsection*{(b) The choice $\boldsymbol{\theta_{\mu \nu} = \theta_{\mu \nu}^{\text{\scriptsize BD}}}$}
Specializing eq.\ \eqref{2.30} we find for the energy and pressure contribution
due to $\theta_{\mu \nu}$:
\begin{align}
\begin{split}
\rho_{\theta} & = \frac{- \vartheta_{0}^{~0}}{8 \pi G}
= - \frac{3}{32 \pi G} \, \left( \frac{\dot G}{G} \right)^{2} ,
\\
p_{\theta} & = \frac{\vartheta_{i}^{~i}}{8 \pi G} =
- \frac{3}{32 \pi G} \, \left( \frac{\dot G}{G} \right)^{2}
\quad \text{($i$ not summed).} 
\end{split}
\label{4.18}
\end{align}
The $00$- and $ii$-components of Einstein's equation are correspondingly
\begin{subequations} \label{4.19}
\begin{gather}
H^{2} + \frac{K}{a^{2}} = 
\frac{1}{3} \, \Lambda + \frac{8 \pi}{3} G \, \rho
+ \left( \frac{\dot G}{G} \right) \, H
- \frac{1}{4} \, \left( \frac{\dot G}{G} \right)^{2},
 \label{4.19a} \\
H^{2} + 2 \, \left( \frac{\ddot a}{a} \right) + \frac{K}{a^{2}}
=
\Lambda - 8 \pi G \, p - \frac{5}{4} \, 
\left( \frac{\dot G}{G} \right)^{2}
+ \frac{\ddot G}{G}+
2 \, \left( \frac{\dot G}{G} \right) \, H.
\label{4.19b}
\end{gather}
\end{subequations}

For the Brans-Dicke choice of the $\theta$-tensor the \textit{on-shell}
 form of the 
consistency condition, eq.\ \eqref{2.23}, reduces to the much simpler residual
condition \eqref{2.38b}. In cosmology, with $T = 3 \, p - \rho$, the latter
boils down to
\begin{align}
4 \pi \, \left( 3 \, p - \rho \right) \, \dot G G = 
\frac{\text{d}}{\text{d} t} \left( G \Lambda \right). \label{4.20}
\end{align}
This condition is equivalent to the original (``off-shell'') version of the
consistency condition, eq.\ \eqref{2.20} or \eqref{2.21}, only when 
\textit{all} field equations are used. Rather than \eqref{4.20} the system
\eqref{4.15} must include the \textit{off-shell} consistency condition \eqref{2.21}
which assumes the form
\begin{align}
\dot \Lambda
+ 8 \pi \, \dot G \, \rho
- \frac{3}{2} \, \left( \frac{\dot G}{G} \right)^{2} \, H
+ 3 \, \left( \frac{\dot G}{G} \right) \, \left( \frac{\ddot a}{a} \right)
+ \frac{3}{2} \, \left( \frac{\dot G}{G} \right)^{3}
- \frac{3}{2} \, \left( \frac{\ddot G \dot G}{G^{2}} \right)
= 0.
\label{4.21}
\end{align}
In deriving \eqref{4.21} from \eqref{2.21} we took 
$D^{\mu} \vartheta_{\mu \nu}^{\text{\scriptsize BD}}$ 
from eq.\ \eqref{2.35}, and we exploited that $G$ is a spatially constant
scalar on which the D'Alembertian $D^{2}$ acts according to
$- D^{2} G = \ddot G + 3 \, H \, \dot G$.

Again it can be checked by a somewhat tedious calculation that \eqref{4.19a}
together with the off-shell consistency condition \eqref{4.21} and the 
continuity equation implies \eqref{4.19b}, provided $a \left( t \right)$
is non-constant.

Up to this point of the discussion the equation of state was kept 
completely arbitrary. For the practical calculations in the following sections
we adopt the linear ansatz
\begin{align}
p \left( \rho \right) = w \, \rho \label{4.22}
\end{align}
where $w$ is a constant. As a result, the continuity equation
\begin{align}
\dot \rho + 3 \, \left( 1+w \right) \, H \, \rho = 0 \label{4.21'}
\end{align}
is easily solved for $\rho$ as a function of $a$:
\begin{align}
\rho \left( t \right) = 
\frac{{\mathcal M}}{8 \pi \, \left[ a \left( t \right) \right]^{3+3w}}
\label{4.22'}.
\end{align}
Here ${\mathcal M}$ is a constant of integration with mass dimension $1-3w$,
defined in the same way as in ref.\ \cite{cosmo1}. With 
\eqref{4.22} the trace of the energy-momentum tensor is
\begin{align}
T = 3 \, p - \rho = \left( 3w-1 \right) \, \rho. \label{4.23}
\end{align}
It vanishes if $w = 1/3$ (``radiation dominance'') or if
${\mathcal M}=0$ (vacuum solutions).
%
%
\section{Solutions of Class \textrm{\textbf{I}}} \label{5}
In this section we discuss various cosmological solutions of Class I.
By definition, $\theta_{\mu \nu} =0$ in this class. For the equation of
state $p = w \, \rho$ the relevant evolution equations are the
modified Friedmann equation \eqref{4.16a} and the off-shell consistency
condition \eqref{4.17} with \eqref{4.22} and \eqref{4.22'} inserted:
\begin{subequations} \label{5.1}
\begin{gather} 
H^{2} + K a^{-2} = 
\frac{1}{3} \Lambda + \frac{1}{3} {\mathcal M} G a^{-3-3w} 
+ H \left( \frac{\dot G}{G} \right),\label{5.1a} 
\\
\dot \Lambda + {\mathcal M} \dot G a^{-3-3w}
+ 3 H \left( \frac{\dot G}{G} \right)^{2} +
3 \left( \frac{\dot G}{G} \right) \left( \frac{\ddot a}{a} \right) = 0.
\label{5.1b}
\end{gather}
\end{subequations}
As these are two independent equations for one function, $a \left( t \right)$,
it is to be expected that the system is not soluble for every 
$G \left( t \right)$, $\Lambda \left( t \right)$. Nevertheless, we shall find 
analytic solutions in the fixed point regime of the
RG flow, as well as for more general power laws $G \propto t^{n}$,
$\Lambda \propto 1 / t^{2}$. Let us discuss these examples in turn.
\subsection{Fixed point solution with $\boldsymbol{K=0}$}
In this subsection we present a power law solution to the eqs.\ \eqref{5.1}
with $G \left( t \right)$ and $\Lambda \left( t \right)$ given by
\eqref{1.50}. This is the time dependence which is expected to occur when the 
underlying RG trajectory is close to a fixed point. The discussion is valid
for a UV- and IR-fixed point alike. In the former case we describe the very
early Universe $ \left( k \to \infty, ~t \to 0 \right)$, while the latter
refers to asymptotically late times $\left( k \to 0, ~t \to \infty \right)$.
To start with let us look at spatially flat Universes, $K=0$.

Inserting a power law ansatz $a \propto t^{\alpha}$ into \eqref{5.1} we
find that indeed \textit{both} equations can be satisfied by a scale factor
of this type provided the constant $\widetilde \lambda_{*}$ assumes the value
\begin{align}
\lambda_{*} \xi^{2} \equiv \widetilde \lambda_{*} =
\frac{2 \, \left( 5-3w \right)}{3 \, \left( 1+w \right)^{2}}.
\label{5.2}
\end{align}
Since $\lambda_{*}$ is fixed once we have picked a specific RG trajectory,
eq.\ \eqref{5.2} is to be interpreted as an equation for $\xi$:
\begin{align}
\xi^{2} = 
\frac{2 \, \left( 5-3w \right)}{3 \, \left( 1+w \right)^{2}}
 \, \frac{1}{\lambda_{*}} \label{5.3}.
\end{align}
The condition \eqref{5.2} and, as a consequence, the
possibility of actually \textit{computing} the factor of proportionality
relating $k$ to $1 / t$, is a direct consequence of the fact that the
system of evolution equations is over-determined; it admits solutions with the
``external field'' $\Lambda \left( t \right) = 
\widetilde \lambda_{*} / t^{2}$ only for one specific value of
$\widetilde \lambda_{*}$.

Henceforth, with an eye towards the UV fixed point found in 
\cite{souma,oliver1,frank1,oliver2,percacciperini}
and the IR fixed point postulated in
\cite{cosmo2} we assume that $g_{*} >0$ and $\lambda_{*} >0$.
As a result, solutions exist only in the range $w < 5/3$ because otherwise
$\xi^{2}$ becomes negative. In the rest of this subsection we make the more
restrictive assumption that $-1 < w < 5/3$.

The $K=0$ fixed point cosmology thus obtained is explicitly given by
\begin{subequations} \label{5.4}
\begin{align}
a \left( t \right) & =
\left[ - \frac{\left( 1+w \right)^{3}}{4 \, \left( 5-3w \right)}
\, {\mathcal M} \, g_{*} \lambda_{*} \right]^{1/ \left( 3+3w \right)}
\, t^{4/ \left( 3+3w \right)}, \label{5.4a} \\
\rho \left( t \right) & =
- \frac{\left( 5-3w \right)}{2 \pi \, \left( 1+w \right)^{3}} \,
\frac{1}{g_{*} \lambda_{*}} \, t^{-4}, \label{5.4b} \\
G \left( t \right) & =
\frac{3 \, \left( 1+w \right)^{2}}{2 \, \left( 5-3w \right)} \,
g_{*} \lambda_{*} \, t^{2}, \label{5.4c} \\
\Lambda \left( t \right) & =
\frac{2 \, \left( 5-3w \right)}{3 \, \left( 1+w \right)^{2}} \,
\frac{1}{g_{*} \lambda_{*}} \, t^{-2}. \label{5.4d}
\end{align}
\end{subequations}
Along with the scale factor we also wrote down the energy density according
to \eqref{4.22} as well as $G \left( t \right)$ and $\Lambda \left( t \right)$
with $\xi$ eliminated everywhere by using \eqref{5.3}. We observe that after
eliminating $\xi$ the observables \eqref{5.4} depend on the fixed point
coordinates $g_{*}$ and $\lambda_{*}$ only via their product. This is a
rather nontrivial and encouraging result because the values of 
$g_{*}$ and $\lambda_{*}$ are scheme dependent, hence unphysical, while
their product $g_{*} \lambda_{*}$ is not 
\cite{percadou,percacciperini,oliver1}. 
Observables
should depend on this product only.

At first sight the solution \eqref{5.4} has a certain similarity with the
cosmology \eqref{A.50} obtained by improving Einstein's equation rather than
the action. However, there is a crucial and, as we shall see, rather 
symptomatic difference: Since $g_{*} \lambda_{*} >0$ and
$w<5/3$, the density \eqref{5.4b} is \textit{negative}. Hence the
conserved quantity
${\mathcal M} \equiv 8 \pi \, \rho \, a^{3+3w}$ is negative, too, and this
is in fact what is needed to make the scale factor \eqref{5.4a} real. 
Clearly a negative density prevents us from interpreting $\rho$
as the energy density due to ordinary (baryonic) matter.

In order to understand this point better we note that for the cosmology
\eqref{5.4} the ``critical'' energy density \eqref{4.11}, which is positive
by definition, is given by
\begin{align}
\rho_{\text{crit}} =
\frac{4 \, \left( 5-3w \right)}{9 \pi \, \left( 1+w \right)^{4}}
\, \frac{1}{g_{*} \lambda_{*}} \, t^{-4}. \label{5.5}
\end{align}
All other densities of interest, $\rho$, $\rho_{\Lambda}$, $\Delta \rho$,
and $\rho_{\theta}$, are proportional to $\rho_{\text{crit}}$,
the constants of proportionality being
\begin{align}
\Omega_{\text{M}} & = - \frac{9}{8} \, \left( 1+w \right) < 0,
& 
\Omega_{\Lambda} & = \frac{1}{8} \, \left( 5-3w \right) > 0, 
\nonumber \\
\Delta \Omega & = \frac{3}{2} \, \left( 1+w \right) > 0,
& 
\Omega_{\theta} & = 0.
\label{5.6}
\end{align}
According to \eqref{4.13}, every $K=0$ cosmology in Class I obeys
\begin{align}
\Omega_{\text{M}} + \Omega_{\Lambda} + \Delta \Omega =1,
\label{5.7}
\end{align}
and clearly the $\Omega$'s of \eqref{5.6} satisfy 
\eqref{5.7}. It is important to note that the new term in \eqref{5.7},
$\Delta \Omega$, which is absent both in standard cosmology and in the 
approach of improving equations, makes always a strictly positive
contribution to the LHS of this equation.
Since $\Omega_{\Lambda}$ cannot become
negative for $\lambda_{*}>0$, the additional positive contribution in
\eqref{5.7} must be compensated by a smaller positive, or even negative value
of $\Omega_{\text{M}}$, as compared to the usual situation where
$\Omega_{\text{M}} + \Omega_{\Lambda}=1$. This is precisely what we found:
$\Omega_{\text{M}}$ turned out negative because it has to counteract a too
strongly positive $\Delta \Omega$-contribution.

The physical interpretation of this phenomenon is as follows. Contrary to the
approach of improving equations, in the present approach of improved
actions the energy and momentum carried by the field $G \left( x \right)$
acts as a source of spacetime curvature. This happens in two different ways:
via the tensor $\Delta T_{\mu \nu}$ which follows straightforwardly from the
variational principle, and via $\theta_{\mu \nu}$. 
In a situation where both approaches
are applicable, and reliable, they should lead to similar results, at least
at a qualitative level. This implies that, roughly
speaking, $\Delta T_{\mu \nu}$ and $\theta_{\mu \nu}$ cancel one another
to some extent. Typically, $\Delta T_{\mu \nu}$ supplies a positive energy 
density, as in the example above, and $\theta_{\mu \nu}$ a negative one.

The prime example is the Brans-Dicke energy-momentum tensor \eqref{2.27} which
(apart from a factor of $3/2$) is the \textit{negative} of the energy-momentum
tensor of an ordinary scalar $\psi$, and correspondingly 
$S_{\theta}^{\text{\scriptsize BD}}$ of \eqref{2.32} contains a kinetic term of the
``wrong'' sign. This becomes explicit in cosmology
where eqs.\ \eqref{4.7a} and \eqref{4.11} lead to
\begin{align}
\Delta \Omega & = \frac{1}{H} \, \left( \frac{\dot G}{G} \right)
\label{5.7'}
\intertext{which is always positive if $\dot G >0$ and $\dot a > 0$, 
whereas eq.\ \eqref{4.18} yields}
\Omega_{\theta}^{\text{\scriptsize BD}} & = 
- \frac{1}{4 \, H^{2}} \, \left( \frac{\dot G}{G} \right)^{2}
\label{5.7''}
\end{align}
which is strictly negative. Similar remarks apply to the pressure.

Above we tried the choice $\theta_{\mu \nu} =0$, and we found that the
corresponding solution of the new approach, eqs.\ \eqref{5.4}, looks quite
different from its counterpart in the old approach, eqs.\ \eqref{A.50}.
The reason is that, as there is no $\theta$-tensor, the positive contributions
from $\Delta T_{\mu \nu}$ must be compensated by negative contributions which
are now forced into $T_{\mu \nu}$ rather than $\theta_{\mu \nu}$.
As a result, we cannot interpret $T_{\mu \nu}$ as the energy-momentum of the
ordinary baryonic matter alone. $T_{\mu \nu}$ is ``contaminated'' by
stress-energy contributions stemming from the fields $G \left( x \right)$
and $\Lambda \left( x \right)$. Therefore the energy density pertaining to
$T_{\mu \nu}$ is to be interpreted as a sum
$\rho = \rho_{\text{mat}} + \rho_{G, \Lambda}$ where $\rho_{\text{mat}}$
is due to the ordinary matter and $\rho_{G, \Lambda}$ to the energy carried
by the fields $G \left( x \right)$ and $\Lambda \left( x \right)$.

In general we have no tool for disentangling $\rho_{\text{mat}}$ from
$\rho_{G, \Lambda}$; the evolution equations determine their sum only.
However, in the light of the discussion above it is a plausible assumption
that $\rho_{G, \Lambda}$ is approximately the negative of $\Delta \rho$.
With $\rho_{G, \Lambda} = -\Delta \rho$ we have 
$\rho = \rho_{\text{mat}} -\Delta \rho$ so that, under this hypothesis, it
is the sum $\rho + \Delta \rho$ which should be identified with the ordinary 
matter energy density.

In order to show that this is actually true we define the
energy-momentum tensor
\begin{align}
\widehat T_{\mu}^{~\nu} & \equiv T_{\mu}^{~\nu} + \Delta T_{\mu}^{~\nu}
\equiv \text{diag} \left[ - \widehat \rho, \widehat p, \widehat p,
\widehat p \, \right]
\label{5.8}
\end{align}
with the entries $\widehat \rho \equiv \rho + \Delta \rho$ and
$\widehat p \equiv p + \Delta p$. Eq.\ \eqref{4.7a} yields for the cosmology
\eqref{5.4}
\begin{align}
\Delta \rho & = 
\frac{2 \, \left( 5-3w \right)}{3 \pi \, \left( 1+w \right)^{3}}
\, \frac{1}{g_{*} \lambda_{*}} \, t^{-4} \label{5.9}
\intertext{and adding the $\rho$ of \eqref{5.4b} leads to}
\widehat \rho \left( t \right) & = - \frac{1}{3} \, \rho \left( t \right).
\label{5.10}
\end{align}
This density is indeed positive and can be identified with $\rho_{\text{mat}}$
therefore.
It will be convenient to define a conserved quantity $\widehat {\mathcal M}$
in terms of $\widehat \rho$ in the same way as ${\mathcal M}$ is defined in
terms of $\rho$, $\widehat {\mathcal M} \equiv 8 \pi \, \widehat \rho
\, a^{3+3w}$, satisfying
\begin{align}
\widehat {\mathcal M} = - \frac{1}{3} \, {\mathcal M} >0
\label{5.11}.
\end{align}
It is instructive to rewrite the solution \eqref{5.4}
in terms of $\widehat \rho$ and $\widehat {\mathcal M}$:
\begin{subequations} \label{5.12}
\begin{align}
a \left( t \right) & =
\left[ \frac{3 \, \left( 1+w \right)^{3}}{4 \, \left( 5-3w \right)}
\, \widehat {\mathcal M} \, g_{*} \lambda_{*} \right]^{1/ \left( 3+3w \right)}
\, t^{4/ \left( 3+3w \right)}, \label{5.12a} \\
\widehat \rho \left( t \right) & =
\frac{\left( 5-3w \right)}{6 \pi \, \left( 1+w \right)^{3}} \,
\frac{1}{g_{*} \lambda_{*}} \, t^{-4}, \label{5.12b} \\
G \left( t \right) & =
\frac{3 \, \left( 1+w \right)^{2}}{2 \, \left( 5-3w \right)} \,
g_{*} \lambda_{*} \, t^{2}, \label{5.12c} \\
\Lambda \left( t \right) & =
\frac{2 \, \left( 5-3w \right)}{3 \, \left( 1+w \right)^{2}} \,
\frac{1}{g_{*} \lambda_{*}} \, t^{-2}. \label{5.12d} \\
\widehat p \left( t \right) & =
\frac{\left( 5-3w \right)}{18 \pi \, \left( 1+w \right)^{3}} \,
\frac{1}{g_{*} \lambda_{*}} \, t^{-4} \label{5.12e}.
\end{align}
\end{subequations}
In the above list we included the pressure $\widehat p = w \, \rho + 
\Delta p$ with
\begin{align}
\Delta p =
\frac{\left( 5-3w \right) \, \left( 1+9w \right)}
{18 \pi \, \left( 1+w \right)^{3}} \,
\frac{1}{g_{*} \lambda_{*}} \, t^{-4}
\label{5.13}
\end{align}
as obtained from \eqref{4.7b} with \eqref{5.4}.

We note in passing that while $p$ and $\rho$ satisfy a simple equation
of state, $p \left( \rho \right) = w \, \rho$, the relationship between
$\Delta p$ and $\Delta \rho$, or $\widehat p$ and $\widehat \rho$, is very
complicated in general, see eqs.\ \eqref{4.7}. However, for the solution 
at hand we
find a remarkably simple and intriguing ``equation of state'' for
$\widehat p$ and $\widehat \rho$:
\begin{align}
\widehat p = \frac{1}{3} \, \widehat \rho
\quad \Longleftrightarrow \quad
\widehat T_{\mu}^{~\mu} =0. \label{5.14}
\end{align}
We emphasize that this relation holds true for \textit{any} value of $w$.
The modified energy-momentum tensor is always traceless, while the
original $T_{\mu \nu}$ is only for $w = 1/3$.

Let us now compare the equations \eqref{5.12}, obtained by improving the 
action, to eqs.\ \eqref{A.50} which resulted from improving the field
equations. Looking at the case $w=1/3$ first, we find that the two cosmologies
are \textit{completely identical} if one identifies the $\rho$ and
${\mathcal M}$ of \eqref{A.50} with the $\widehat \rho$ and
$\widehat {\mathcal M}$ in \eqref{5.12}. This confirms our hypothesis that in 
the Class I-solutions it is the sum $\rho + \Delta \rho$ which is to be
identified with $\rho_{\text{mat}}$, while in the framework of improved 
equations it is $\rho$ itself. Even the result for the pressure is the same
in both cases since, by \eqref{5.14}, we have $\widehat p = \widehat \rho \, /
\, 3$ for any $w$, and this equation of state accidentally coincides with 
$p = w \, \rho$ if $w = 1/3$. 

Also for other values of $w$ in the interval $\left( -1, 5/3 \right)$ the
two cosmologies are qualitatively similar; the time dependencies of all 
quantities
of interest are the same, only the prefactors of the various powers of $t$
differ slightly. Let us denote the functions \eqref{A.50} where ${\mathcal M}$
is replaced with $\widehat {\mathcal M}$ by $a_{\text{ieq}}$,
$\rho_{\text{ieq}}$, $G_{\text{ieq}}$, and $\Lambda_{\text{ieq}}$,
respectively, with ``ieq'' standing for ``improved equation''. Comparing them
to their analogs in \eqref{5.12} we find the time-independent ratios
\begin{subequations} \label{5.15}
\begin{align}
\frac{a \left( t \right)}{a_{\text{ieq}} \left( t \right)} & =
\left[ \frac{3}{16} \, \left( 1+w \right) \, \left( 5-3w \right) 
\right]^{-1 / \left( 3+3w \right)} \label{5.15a} \\
\frac{\widehat \rho \left( t \right)}{\rho_{\text{ieq}} \left( t \right)} & =
\frac{3}{16} \, \left( 1+w \right) \, \left( 5-3w \right)
\label{5.15b} \\
\frac{G \left( t \right)}{G_{\text{ieq}} \left( t \right)} & =
\frac{4}{5-3w} \label{5.15c} \\
\frac{\Lambda \left( t \right)}{\Lambda_{\text{ieq}} \left( t \right)} & =
\frac{1}{4} \, \left( 5-3w \right). \label{5.15d}
\end{align}
\end{subequations}
For $w=1/3$ all ratios are exactly equal to $1$, and for $w$ arbitrary but not 
too close to the boundaries of the interval $\left( -1, 5/3 \right)$
they are still rather close to unity. For $w=0$, say, one has $0.98$, $0.94$,
$0.80$, and $1.25$, respectively.

Thus we may conclude that with the reinterpretation of $\rho_{\text{mat}}$
as $\rho + \Delta \rho$ the two  approaches, improving equations and 
improving actions, lead to very similar results, the ratios \eqref{5.15}
being a measure of their quantitative precision.

The only minor difference concerns the pressure. The
corresponding ratio is
\begin{align}
\frac{\widehat p \left( t \right)}{p_{\text{ieq}} \left( t \right)} & =
\frac{\left( 1+w \right) \, \left( 5-3w \right)}{16w}
\qquad \left( w \neq 0 \right) \label{5.16}
\end{align}
It can become large when $w$ is close to zero. In fact, for $w=0$ we have
$p_{\text{ieq}} =0$, but
\begin{align}
\widehat p \left( t \right) = \Delta p \left( t \right) 
= \frac{5}{18 \pi} \, \frac{1}{g_{*} \lambda_{*}} \, t^{-4}
\label{5.17}
\end{align}
is nonzero.
\subsection{Fixed point solutions with $\boldsymbol{K= \pm 1}$}
In the case of a spatially curved Universe, too, it is possible to find
solutions of Class I with $G \left( t \right)$ and $\Lambda \left( t \right)$
evolving according to the fixed point law \eqref{1.50}, albeit only for the 
equation of state with
\begin{align}
w = 1/3. \label{5.18}
\end{align}
From eqs.\ \eqref{5.1} with $K=+1$ or $-1$ we obtain the following 
cosmologies with a linearly growing scale factor:
\begin{subequations} \label{5.19}
\begin{align}
a \left( t \right) & =
\left[ \frac{1}{3} \,
\left( 
-K + 2 \, \sqrt{1 - \tfrac{1}{3} \, {\mathcal M} \, g_{*} \lambda_{*}\,} \,
\right)
\right]^{1/2} \, t \label{5.19a} \\
\rho \left( t \right) & =
\frac{9}{8 \pi} \, \frac{{\mathcal M}}
{\left( 
-K + 2 \, \sqrt{1 - \tfrac{1}{3} \, {\mathcal M} \, g_{*} \lambda_{*}\,}\,
\right)^{2}} \, t^{-4} \label{5.19b} \\
G \left( t \right) & =
\frac{1}{3} \,
\frac{\left( 
-K + 2 \, \sqrt{1 - \tfrac{1}{3} \, {\mathcal M} \, g_{*} \lambda_{*}\,}\,
\right)}
{\left( 
K + \sqrt{1 - \tfrac{1}{3} \, {\mathcal M} \, g_{*} \lambda_{*}\,}\,
\right)} \, g_{*} \lambda_{*} \, t^{2} \label{5.19c} \\
\Lambda \left( t \right) & =
3 \, \frac{\left( 
K + \sqrt{1 - \tfrac{1}{3} \, {\mathcal M} \, g_{*} \lambda_{*}\,}\,
\right)}
{\left( 
-K + 2 \, \sqrt{1 - \tfrac{1}{3} \, {\mathcal M} \, g_{*} \lambda_{*}\,}\,
\right)}
\, \frac{1}{g_{*} \lambda_{*}} \, t^{-2}. \label{5.19d}
\end{align}
\end{subequations}
As in the case $K=0$, the evolution equations fix the value of 
$\widetilde \lambda_{*}$. In writing down \eqref{5.19} we used this
information in order to express $\xi$ in terms of $\lambda_{*}$
everywhere. They are related by
\begin{align}
\xi^{2} = \frac{3}{2} \, \left[ 1 + \frac{K}{A^{2}} \right] \, 
\frac{1}{\lambda_{*}} \label{5.20}
\end{align}
where $A^{2} \equiv \frac{1}{3} \, \left[-K + 2 \, 
\sqrt{1 - {\mathcal M} \, g_{*} \lambda_{*} / 3\,} \,\right]$.
To make sense, $\xi$ must be real, and this condition leads to a constraint
on the ``matter'' contents of the Universe as parameterized by the constant
${\mathcal M}$. Assuming, as always, $g_{*}>0$ and $\lambda_{*}>0$, it reads
\begin{align}
\begin{split}
{\mathcal M} & < \frac{9}{4 \, g_{*} \lambda_{*}} \phantom{0}
 \quad \text{ for } K=+1 \\
{\mathcal M} & < 0 \phantom{\frac{9}{4 \, g_{*} \lambda_{*}}}
 \quad \text{ for } K=-1. 
\end{split} \label{5.21}
\end{align}
If \eqref{5.21} is satisfied, the scale factor as well as $G$ and $\Lambda$
are real and positive in \eqref{5.19}. For $K=-1$ the situation is similar as
in the previous subsection: the energy density $\rho$ is negative for all 
allowed values of ${\mathcal M}$. A new phenomenon is encountered in the 
spherical case $K=+1$. Here there exists a ``window'' of ${\mathcal M}$-values
between zero and $9 / \left( 4 \, g_{*} \lambda_{*} \right)$ for
which the density $\rho$ is positive.

Again, the various energy densities are all proportional to the critical
energy density,
\begin{align}
\rho_{\text{crit}} & =
\frac{9}{8 \pi}
\frac{\left( 
K + \sqrt{1 - \frac{1}{3} \, {\mathcal M} \, g_{*} \lambda_{*}\,}\,
\right)}
{\left( 
-K + 2 \, \sqrt{1 - \frac{1}{3} \, {\mathcal M} \, g_{*} \lambda_{*}\,}\,
\right)}
\,
\frac{1}{g_{*} \lambda_{*}} \, t^{-4} \label{5.22},
\end{align}
but the corresponding $\Omega$'s depend on ${\mathcal M}$ now:
\begin{align}
\begin{split}
\Omega_{\text{M}} & =
\frac{{\mathcal M} \, g_{*} \lambda_{*}}
{\left( 
K + \sqrt{1 - \frac{1}{3} \, {\mathcal M} \, g_{*} \lambda_{*}\,}\,
\right) \,
\left( 
-K + 2 \, \sqrt{1 - \frac{1}{3} \, {\mathcal M} \, g_{*} \lambda_{*}\,}\,
\right)}, \\
\Omega_{\Lambda} & =
\frac{\left( 
K + \sqrt{1 - \frac{1}{3} \, {\mathcal M} \, g_{*} \lambda_{*}\,}\,
\right)}
{\left( 
-K + 2 \, \sqrt{1 - \frac{1}{3} \, {\mathcal M} \, g_{*} \lambda_{*}\,}\,
\right)}, \\
\Delta \Omega & = 2, \quad \Omega_{\theta} = 0.
\end{split}
\end{align}
Note that $\Delta \Omega$ has the same positive value as for $K=0$ with
$w=1/3$.
\paragraph{A Vacuum Solution:}
Picking $K=+1$, the cosmology \eqref{5.19} has a well-behaved limit\footnote{
With $K=-1$, Newton's constant \eqref{5.19c} diverges in the limit 
${\mathcal M} \to 0$. Also the $K=0$ cosmology \eqref{5.4} becomes singular
in this limit, its scale factor $a \left( t \right)$ vanishes identically.}
${\mathcal M} \to 0$. It describes a linearly expanding Universe with
spherical
time slices which does not contain any real matter; the expansion is driven
by $\Delta T_{\mu \nu}$ and the cosmological constant alone.
Apart from the universal product $g_{*} \lambda_{*}$, this cosmology does not
involve any free parameter:
\begin{subequations} \label{5.24}
\begin{align}
a \left( t \right) & = \frac{1}{\sqrt{3}} \, t, \label{5.24a} \\
\rho \left( t \right) & = 0, \label{5.24b} \\
G \left( t \right) & = \frac{1}{6} \, g_{*} \lambda_{*} \, t^{2},
\label{5.24c} \\
\Lambda \left( t \right) & = 6 \, t^{-2}. \label{5.24d}
\end{align}
\end{subequations}
It is easily checked explicitly that this limiting case satisfies all relevant
evolution equations. The parameter $\xi$ is fixed by $\xi^{2} = 6 /
\lambda_{*}$, and the densities are $\Omega_{\text{M}}=0$, 
$\Omega_{\Lambda}=2$, $\Delta \Omega=2$, $\Omega_{\theta}=0$.

This vacuum solution owes its existence to the tensor $\Delta T_{\mu \nu}$;
improving the field equation rather than the action one finds no analogous
solution. This is a further indication that in the vacuum sector
the two approaches can yield
similar results only when an appropriate $\theta$-tensor is included.
\subsection{General power laws $\boldsymbol{\left( K=0 \right)}$} \label{4.C}
Being over-determined, the system of equations
\eqref{5.1} restricts the allowed backgrounds
$\left( G \left( t \right), \Lambda \left( t \right) \right)$. 
This feature reduces ambiguities related
to the non-universal properties of the RG flow and to the mathematical 
modeling of the physical cutoff mechanism by the identification 
$k=k \left( x \right)$. In the fixed point regime the restrictions were 
rather mild, only the parameter $\widetilde \lambda_{*}$ got fixed. We shall
now investigate backgrounds of the type
\begin{gather}
G \left( t \right) = C \, t^{n}, \quad \Lambda \left( t \right) = D \, t^{-m}
\label{5.25}
\end{gather}
where $C>0$, $D$, $n$, and $m$ are a priori arbitrary real constants.
Here the restrictions on allowed backgrounds 
are seen much more clearly. The only power law
solutions admitted by \eqref{5.1}, with $K=0$, $w \geq -1$, and
${\mathcal M} \neq 0$, are the following three families.
\subsubsection*{1. First family $\boldsymbol{\Bigl( n \neq -2,~w >-1 \Bigr)}$}
The solutions belonging to this family are labeled by an arbitrary real
exponent $n$, different from $-2$, and by the parameters $C$ and
${\mathcal M}$. Imposing solubility of the system, $m$ and $D$ are uniquely
determined:
\begin{subequations} \label{5.26}
\begin{align}
a \left( t \right) & =
\left[ \frac{3 \, \left( 1+w \right)^{2}}
{4 - 3 \, n \, \left( 1+w \right) - n^{2} \, \left( 4 + 3w \right)}
\, {\mathcal M} C \right]^{1 / \left( 3+3w \right)}
\, t^{\left( n+2 \right) / \left( 3+3w \right)}, \label{5.26a} \\
\rho \left( t \right) & =
\frac{1}{24 \pi} \,
\frac{\bigl[
4 - 3 \, n \, \left( 1+w \right) - n^{2} \, \left( 4 + 3w \right) \bigr]}
{\left( 1+w \right)^{2}} \, \frac{1}{C} \, t^{- \left( n+2 \right)},
\label{5.26b} \\
G \left( t \right) & = C \, t^{n}, \label{5.26c} \\
\Lambda \left( t \right) & =
\frac{n}{3} \, \frac{\left( 2 \, n + 1 - 3w \right)}
{\left( 1+w \right)^{2}} \, t^{-2}. \label{5.26d}
\end{align}
\end{subequations}
For the special choice $n=+2$, $C= \frac{3}{2} \, \left( 1+w \right)^{2}
\, g_{*} \lambda_{*} / \left( 5-3w \right)$, eqs.\ \eqref{5.26} reproduce
the fixed point solution \eqref{5.4}. A new feature of \eqref{5.26} is that
if $n$ lies in a narrow interval $\left[ n_{-}, n_{+} \right]$ there
exist solutions with positive energy density (${\mathcal M}>0$).
These solutions are more the exception than the rule, however; for all
$n<n_{-}$ and $n>_{+}$ the cosmology \eqref{5.26} exists only if
${\mathcal M}<0$. (Here we assume $C>0$ so that $G$ is positive.) The limits
of the interval depend on the equation of state:
\begin{align}
n_{\pm} & = - \frac{3}{2} \, \left( \frac{1+w}{4+3w} \right)
\pm \left[ \frac{4}{\left( 4+3w \right)} + \frac{9}{4} \,
\left( \frac{1+w}{4+3w} \right)^{2} \right]^{1/2} \label{5.27}
\end{align}
For dust and radiation, say,
\begin{align}
\begin{split}
n_{-} \approx -1.44,~n_{+} \approx 0.69 \qquad& (w=0) \\
n_{-} \approx -1.38,~n_{+} \approx 0.58 \qquad & (w=1/3)
\end{split}
\label{5.28}
\end{align}
so that the window for positive energy solutions is indeed
comparatively small.
\subsubsection*{2. Second family $\boldsymbol{\Bigl( n=-2,~w>-1 \Bigr)}$}
There exists an exceptional cosmology where both $\Lambda$ and $G$
decay $\propto t^{-2}$. Assuming, as always, that $C>0$ it exists only
for negative ${\mathcal M}$ and $\rho$:
\begin{subequations} \label{5.29}
\begin{align}
a \left( t \right) & =
\left[ - \frac{1}{2} \, \left( 1+w \right) \, {\mathcal M} C \right]^{1 /
  \left( 3+3w \right)}=const, \label{5.29a} \\
\rho \left( t \right) & =
- \frac{1}{4 \pi} \, \frac{1}{\left( 1+w \right)} \, \frac{1}{C}=const,
\label{5.29b} \\
G \left( t \right) & = C \, t^{-2}, \label{5.29c} \\
\Lambda \left( t \right) & = \frac{2}{\left( 1+w \right)} \, t^{-2}.
\label{5.29d}
\end{align}
\end{subequations}
This cosmology is quite exotic in that $G$ and $\Lambda$ depend on time but
the Universe does not expand.\footnote{Since $\dot a =0$, the $ii$-component
of Einstein's equation must be checked explicitly; it is indeed found to be
satisfied by \eqref{5.29}.} Even though the scale factor is constant, this
Universe has an initial singularity (``big bang'') at which $G$ and
$\Lambda$ diverge.
\subsubsection*{3. Third family $\boldsymbol{\Bigl( w=-1 \Bigr)}$}
For the equation of state with $w=-1$ there exists another exotic
cosmology with a time independent density, and with $a$, $G$, and
$\Lambda$ increasing proportional to $\sqrt{t}$:
\begin{subequations} \label{5.30}
\begin{align}
a \left( t \right) & =
A \, t^{1/2}, \label{5.30a} \\
\rho \left( t \right) & =
\frac{{\mathcal M}}{8 \pi},
\label{5.30b} \\
G \left( t \right) & = C \, t^{1/2}, \label{5.30c} \\
\Lambda \left( t \right) & = - {\mathcal M} C \, t^{1/2}.
\label{5.30d}
\end{align}
\end{subequations}
The normalization of the scale factor, $A$, is completely arbitrary.
%
%
\section{Solutions of Class \textrm{\textbf{II}}} \label{6}
According to Subsection \ref{2.D}, the Class II is defined by
$\theta_{\mu \nu} = \theta_{\mu \nu}^{\text{\scriptsize BD}}$, 
$\Lambda =0$, and $T=0$ which translates into $w = 1/3$ or ${\mathcal M}=0$
in the present setting.
The relevant cosmological evolution equations are the modified
Friedmann equation \eqref{4.19a} and the off-shell consistency condition
\eqref{4.21} with $\Lambda \equiv 0$, $w=1/3$, and $\rho = {\mathcal M} /
\left( 8 \pi \, a^{4} \right)$ inserted:
\begin{subequations} \label{6.1}
\begin{gather}
H^{2} + \frac{K}{a^{2}} = \frac{{\mathcal M}}{3} \, G \, a^{-4}
+ \left( \frac{\dot G}{G} \right) \, H
- \frac{1}{4} \, \left( \frac{\dot G}{G} \right)^{2},
\label{6.1a} \\
{\mathcal M} \, \dot G \, a^{-4}
- \frac{3}{2} \, \left( \frac{\dot G}{G} \right)^{2} \, H
+ 3 \, \left( \frac{\dot G}{G} \right) \, \left( \frac{\ddot a}{a} \right)
+ \frac{3}{2} \, \left( \frac{\dot G}{G} \right)^{3}
- \frac{3}{2} \, \left( \frac{\ddot G \dot G}{G^{2}} \right)
= 0
\label{6.1b}
\end{gather}
\end{subequations}
Vacuum solutions are obtained from \eqref{6.1} with ${\mathcal M}=0$.
\subsection{Generating solutions via Weyl transformations}
According to the discussion in Subsection \ref{2.E} it should be possible to
obtain the solutions to the system \eqref{6.1} by Weyl-transforming the
solutions of the much simpler system \eqref{2.52} in which Newton's constant
is really constant. There arises the following problem, however.
Assume we have solved the simpler system and obtained a line element
$\text{d} s_{\gamma}^{2} \equiv \gamma_{\mu \nu} \text{d} x^{\mu} 
\text{d} x^{\nu}$ which has the standard Robertson-Walker form \eqref{4.1}.
According to \eqref{2.53a} the line element we are actually after,
$\text{d} s^{2} \equiv g_{\mu \nu} \text{d} x^{\mu} \text{d} x^{\nu}$,
obtains as
\begin{align}
\text{d} s^{2} = \frac{G \left( t \right)}{\, \overline{G} \,} \,
\text{d} s_{\gamma}^{2}, \label{6.2}
\end{align}
and this is not of the Robertson-Walker form. As we shall discuss
next, this defect can be repaired by a reparametrization of the time 
coordinate.

Let us write $\text{d} s_{\gamma}^{2}$ in the style of \eqref{4.1},
\begin{align}
\text{d} s_{\gamma}^{2} = - \text{d} \tau^{2} +
b^{2} \left( \tau \right) \, \text{d} \Omega_{K}^{2} \label{6.3}
\end{align}
with the cosmological time $\tau$ and the scale factor $b \left( \tau \right)$.
Inserting \eqref{6.3} into the constant-$G$ Einstein equation \eqref{2.52a}
leads to the classical equations
\begin{subequations} \label{6.4}
\begin{gather}
H_{b}^{2} + \frac{K}{b^{2}} = \frac{8 \pi \overline{G}}{3} \, \check \rho
\label{6.4a} \\
H_{b}^{2} + \frac{2}{b} \, \frac{\text{d}^{2} b}{\text{d} \tau^{2}}
+ \frac{K}{b^{2}} = - \frac{8 \pi \overline{G}}{3} \, \check \rho
\label{6.4b}
\end{gather}
\end{subequations}
with
$H_{b} \equiv b^{-1} \, \left( \text{d} b / \text{d} \tau \right)$.
We wrote $\check \rho$ and $\check \rho / 3$ for the density and
pressure corresponding to the energy-momentum tensor in the transformed
frame, $T_{\mu \nu} \left( {\mathcal A, \gamma} \right)$. The substitute for the
${\mathcal A}$-equation of motion is the continuity equation
\begin{gather}
\frac{\text{d}}{\text{d} \tau} \, \check \rho \left( \tau \right) 
+ 4 \, H_{b} \left( \tau \right) \, \check \rho \left( \tau \right) =0
\label{6.5'}
\end{gather}
which integrates to
$\check \rho = {\mathcal M} / \bigl( 8 \pi \, b^{4} \bigr)$
whence the $00$- and $ii$-components of the constant-$G$ equation become
\begin{subequations} \label{6.7}
\begin{gather}
H_{b}^{2} + \frac{K}{b^{2}} 
= \frac{1}{3} \, {\mathcal M} \overline{G} \, b^{-4}
\label{6.7a} \\
H_{b}^{2} + \frac{2}{b} \, \frac{\text{d}^{2} b}{\text{d} \tau^{2}}
+ \frac{K}{b^{2}} = - \frac{1}{3} \, {\mathcal M} \overline{G} \, b^{-4}
\label{6.7b}. 
\end{gather}
\end{subequations}
We know that \eqref{6.4a} with \eqref{6.5'} implies \eqref{6.4b}, and that
\eqref{6.7a} implies \eqref{6.7b} if $b \neq const$. We adopt \eqref{6.7a}
as the (only) independent $b$-equation.

Let us pick a solution $b \left( \tau \right)$. Then, in the 
$\left( \tau, r, \theta, \varphi \right)$-coordinate system, the metric
$g_{\mu \nu}$ is represented by
\begin{align}
\begin{split}
\text{d} s^{2}
& = \left[ G / \, \overline{G} \, \right] \,
\left( - \text{d} \tau^{2} + b^{2} \left( \tau \right) \,
\text{d} \Omega_{K}^{2} \right)
\\
& = - \left( \sqrt{G / \, \overline{G} \,} ~\text{d} \tau \right)^{2}
+ \left( \sqrt{G / \, \overline{G} \,} ~ b \left( \tau \right) \right)^{2} \,
 \text{d} \Omega_{K}^{2}
\end{split} \label{6.8}
\end{align}
where $G$ is considered a function of $\tau$ a priori.
Now we introduce a new time coordinate $t = t \left( \tau \right)$ such
that \eqref{6.8} assumes the standard form 
$\text{d} s^{2} = - \text{d} t^{2} + a^{2} \left( t \right) \text{d} \Omega_{K}^{2}$.
Obviously we need that $\text{d} t = \sqrt{G / \overline{G} \,} ~\text{d}
\tau$, and since we would like to prescribe $G$ in the final $t$- rather than
the original $\tau$-coordinate system this condition provides us with the
derivative of $\tau = \tau \left( t \right)$,
\begin{align}
\frac{\text{d}}{\text{d} t} \, \tau \left( t \right) & =
\sqrt{\, \overline{G} / G \left( t \right) \,}, \label{6.9}
\end{align}
which can be integrated immediately,
\begin{subequations} \label{6.10}
\begin{align}
\tau \left( t \right) & =
\int_{t_{1}}^{t} \text{d} t^{\prime} ~
\sqrt{\, \overline{G} / G \left( t^{\prime} \right) \,}
\label{6.10a},
\intertext{with a constant $t_{1}$. 
The final result for the scale factor $a$ is obtained by a 
combined Weyl and general coordinate transformation:}
a \left( t \right) & =
\sqrt{G \left( t \right) / \, \overline{G} \,} ~ b \left( \tau \left( t 
\right) \right). \label{6.10b}
\end{align}
\end{subequations}

The equations \eqref{6.10a} and \eqref{6.10b} express the ``magic'' of
the cosmological Class II-solutions. Rather than dealing with the 
complicated-looking system \eqref{6.1} directly it is 
sufficient to solve the
standard Friedmann equation of the radiation dominated Universe, 
eq.\ \eqref{6.7a}. Because of this hidden simplicity, the system \eqref{6.1}
is not over-determined, as was its analog for $\theta_{\mu \nu} =0$.
It is soluble for arbitrary prescribed functions $G \left( t \right)$.

When one inserts \eqref{6.10} into \eqref{6.1a} and \eqref{6.1b}
it is impressive
 to see explicitly that both of those rather complicated
equations are satisfied identically if $b \left( \tau
\right)$ solves \eqref{6.7a}. The calculation makes essential use of the 
following relations between the first and second derivatives
of $a \left( t \right)$ and $b \left( \tau \right)$:
\begin{gather}
H = \sqrt{\, \overline{G} / G \left( t \right) \,} ~ H_{b}
+ \frac{1}{2} \, \left( \frac{\dot G}{G} \right) \label{6.11}
\\
\frac{\ddot a}{a} = 
\left[ \, \overline{G} / G \left( t \right) \right] \, 
\left( \frac{1}{b} \, \frac{\text{d}^{2} b}{\text{d} \tau^{2}} \right)
- \frac{1}{2} \, \left( \frac{\dot G}{G} \right)^{2}
+ \frac{1}{2} \, \left( \frac{\ddot G}{G} \right) 
+ \frac{1}{2} \, H \, \left( \frac{\dot G}{G} \right)
\label{6.12}
\end{gather}
(As usual, the dot denotes the derivative with respect to $t$.)

In checking the $a$-equations it becomes obvious that the constant of
integration ${\mathcal M}$ must have the same value in the $b$- and the
$a$-system of equations. 
In fact, only then the respective energy densities
are related by
\begin{align}
\rho \left( t \right) =
\left[ G \left( t \right) / \, \overline{G} \, \right]^{-2} \,
\check \rho \left( \tau \left( t \right) \right) \label{6.13}
\end{align}
so that $\rho$ transforms according to \eqref{2.53b} with the expected
Weyl weight $\Delta_{\rho} = 2$.
\subsection{Vacuum spacetimes}
The simplest situation with $T=0$ is the absence of any matter so that
$T_{\mu \nu} =0$ and $\rho$, $\check \rho$, ${\mathcal M}$ $=0$.
The relevant solutions to the $b$-equations \eqref{6.4} are
Minkowski space,
\begin{gather}
b \left( \tau \right) = b_{0} = const, \quad K=0, \label{6.14}
\intertext{and the Milne Universe which is merely an unconventional 
coordinatization of flat spacetime with hyperbolic time slices:}
b \left( \tau \right) = \tau, \quad K=-1. \label{6.15}
\end{gather}

According to the discussion of the previous subsection, Minkowski space gives
rise to an improved cosmology with
\begin{align}
a \left( t \right) = b_{0} \, \sqrt{G \left( t \right) / \, \overline{G} \,},
\quad K=0, \label{6.16}
\end{align}
and $\rho \equiv 0$, $\Lambda =0$. Here $G \left( t \right)$ can be any
function of time. The corresponding Hubble constant is
$H= \dot G / \left( 2 \, G \right)$, whence, with \eqref{5.7'} and 
\eqref{5.7''},
\begin{gather}
\Omega_{\text{M}} =0, \quad \Omega_{\Lambda} =0, \quad
\Delta \Omega =2, \quad \Omega_{\theta}=-1. \label{6.17}
\end{gather}
We observe that the density contribution of 
$\theta_{\mu \nu}^{\text{\scriptsize BD}}$ indeed counteracts the one from
$\Delta T_{\mu \nu}$, without completely canceling it though. Interestingly
enough, also in this class of cosmologies the distinguished time dependence
$G \propto t^{2}$ leads to a linear expansion of the Universe, 
$a \propto t$.

In an analogous fashion the Milne Universe generates the following vacuum
solution, with zero cosmological constant and hyperbolic 3-space:
\begin{align}
a \left( t \right) = \int_{t_{1}}^{t} \! \! \text{d} t^{\prime} ~
\sqrt{G \left( t \right) / G \left( t^{\prime} \right)}, \quad K=-1.
\label{6.18}
\end{align}
For $G \propto t^{2}$, say, this scale factor behaves as $a \propto t \,
\ln t$.

There are no further vacuum solutions of Class II beyond \eqref{6.16}
and \eqref{6.18}, in particular there are none for $K=+1$. This can be seen 
directly from the Friedmann equation \eqref{6.1a} which, 
provided ${\mathcal M}=0$, can be cast into the remarkable form
\begin{align}
\left[ \dot a - \frac{1}{2} \left( \frac{\dot G}{G} \right) \, a
\right]^{2} = -K \label{6.19}.
\end{align}
Obviously \eqref{6.19} cannot be satisfied for $K=+1$, while for
$K=0$ and $K=-1$, respectively, its most general solutions 
(with $a>0$) are given by \eqref{6.16} and \eqref{6.18}.
\subsection{A radiation dominated Universe}
The simplest example of a non-vacuum spacetime with $T=0$ is induced
by the familiar spatially flat radiation dominated Universe with zero 
cosmological constant:
\begin{align}
b \left( \tau \right) & =
\left[ \frac{4}{3} \, {\mathcal M} \overline{G} \right]^{1/4}
\, \tau^{1/2}, \quad K=0. \label{6.20}
\intertext{Its Weyl-transform describes a $K=0$-cosmology with scale
factor}
a \left( t \right) & =
\left[ \frac{4}{3} \, {\mathcal M} \right]^{1/4} \,
\left( G \left( t \right) \, \int_{t_{1}}^{t} \! \! 
\frac{\text{d} t^{\prime}}{\sqrt{G \left( t^{\prime} \right)}}
\right)^{1/2}, \label{6.21}
\end{align}
density $\rho= {\mathcal M} / \left( 8 \pi \, a^{4} \right)$,
pressure $p=\rho / 3$, and $\Lambda \left( t \right)=0$. 

The important
point to be noted here is that this solution exits if, and only if,
${\mathcal M}>0$.
Thus the Universe is filled with matter of positive energy density $\rho$.
Therefore, contrary to the $\theta_{\mu \nu}=0$-case discussed in the previous
section, we may now interpret $\rho$ as the energy density of ordinary
baryonic matter. In accordance with our earlier discussion, the Brans-Dicke
$\theta$-tensor included here compensates for certain contributions of 
$\Delta T_{\mu \nu}$, which then no longer ``contaminate'' $\rho$ and
$p$ with contributions which are actually due to the position-dependence
of $G$.
%
%
\section{Solutions of Class \textrm{\textbf{III}}} \label{7}
In Class III, $\theta_{\mu \nu}=\theta_{\mu \nu}^{\text{\scriptsize BD}}$ and
$\Lambda \neq 0$. The corresponding evolution equations, the $00$-Einstein
equation and the off-shell consistency condition, read, respectively,
\begin{subequations}\label{7.1}
\begin{gather}
H^{2} + \frac{K}{a^{2}} = \frac{1}{3} \, \Lambda
+ \frac{1}{3} \, {\mathcal M} G \, a^{-3-3w}
+ H \, \left( \frac{\dot G}{G} \right)
- \frac{1}{4}  \left( \frac{\dot G}{G} \right)^{2}, \label{7.1a}
\\
\dot \Lambda + {\mathcal M} \dot G \, a^{-3-3w}
- \frac{3}{2} \, H \, \left( \frac{\dot G}{G} \right)^{2}
+ 3 \, \left( \frac{\ddot a}{a} \right) \, \left( \frac{\dot G}{G} \right) 
+ \frac{3}{2} \, \left( \frac{\dot G}{G} \right)^{3}
- \frac{3}{2} \left( \frac{\ddot G \dot G}{G^{2}} \right) =0
\label{7.1b}
\end{gather}
\end{subequations}
Generically this system of equations cannot be solved by the 
Weyl-transformation technique. An exception are the solutions in the 
sub-class IIIa
to which we turn first.
\subsection{The Class \textrm{\textbf{IIIa}}}
According to our earlier definition, the Class IIIa is characterized by
$G \Lambda = const$ and $T=0$, i.\,e.\ by
\begin{gather}
\Lambda \left( t \right) = \frac{\overline{G} \, \overline{\Lambda}}
{G \left( t \right)}, 
\quad \text{and} \quad
w = \frac{1}{3}
\quad \text{or} \quad
{\mathcal M}=0. \label{7.2}
\end{gather}
Very much as in the Class II, solutions of this type can be obtained by
Weyl-transforming solutions of the corresponding constant-$G$,
constant-$\Lambda$ Einstein equation \eqref{2.57a}, this time with
$\overline{\Lambda} \neq 0$ though. As a result, the $b \left( \tau \right)$-system
reads
\begin{subequations} \label{7.3}
\begin{gather}
H_{b}^{2} + \frac{K}{b^{2}} =
\frac{1}{3} \, \overline{\Lambda} 
+ \frac{1}{3} \, {\mathcal M} \overline{G} \, b^{-4}, \label{7.3a} \\
H_{b}^{2} + \frac{2}{b} \, \frac{\text{d}^{2} b}{\text{d} \tau^{2}} \, 
+ \frac{K}{b^{2}} = \overline{\Lambda}
- \frac{1}{3} \, {\mathcal M} \overline{G} \, b^{-4}. \label{7.3b}
\end{gather}
\end{subequations}
Here we put $w=1/3$; to get the corresponding vacuum equations one sets
${\mathcal M}=0$. Every solution $b \left( \tau \right)$ of \eqref{7.3}
induces a solution $a \left( t \right)$ of \eqref{7.1} with \eqref{7.2}.
As before, it is given by \eqref{6.10b} with \eqref{6.10a}. The 
construction works for any given function $G \left( t \right) 
\propto 1 / \Lambda \left( t \right)$.

The Class IIIa includes the fixed point regime where $G = \widetilde g_{*}
\, t^{2}$, $\Lambda = \widetilde \lambda_{*} \, t^{-2}$, and
$\overline{G} \, \overline{\Lambda} =  \widetilde g_{*} \widetilde
\lambda_{*}$. For this time dependence, the relationship between $t$ and
$\tau$ is given by
\begin{align}
\tau \left( t \right) =
\sqrt{\,\overline{G} / \widetilde g_{*} \,} ~ \ln \left( t / t_{1} \right)
\label{7.4}.
\end{align}

Next we look at some instructive examples.
\subsubsection{Vacuum solutions from (anti) de Sitter space}
For ${\mathcal M} =0$, the $b$-system \eqref{7.3} has the following well-known
solutions which describe (a part of) de Sitter space ($\overline{\Lambda}
>0$) or anti-de Sitter space ($\overline{\Lambda}<0$), respectively:
\begin{subequations} \label{7.5}
\begin{align}
b \left( \tau \right) & =
b_{0} \, \exp \left( \pm \sqrt{\, \overline{\Lambda} / 3 \,} ~ \tau \right)
& & (K=0,~ \overline{\Lambda}>0) \label{7.5a} \\
b \left( \tau \right) & =
\sqrt{3 / \, \overline{\Lambda}\, } ~
\cosh \left( \sqrt{\, \overline{\Lambda} / 3 \,} ~ \tau \right)
& & (K=+1,~ \overline{\Lambda}>0) \label{7.5b} \\
b \left( \tau \right) & =
\sqrt{3 / \, \overline{\Lambda} \,} ~ 
\sinh \left( \sqrt{\, \overline{\Lambda} / 3 \,} ~ \tau \right)
& & (K=-1,~ \overline{\Lambda}>0) \label{7.5c} \\
b \left( \tau \right) & =
\sqrt{-3 / \, \overline{\Lambda}} ~ 
\cos \left( \sqrt{-\overline{\Lambda} / 3 \,} ~ \tau \right)
& & (K=-1,~ \overline{\Lambda}<0) \label{7.5d}.
\end{align}
\end{subequations}
Restricting ourselves to the fixed point regime, these scale factors imply
the following solutions to the original system \eqref{7.1} with
${\mathcal M} =0$:
\begin{subequations} \label{7.6}
\begin{align}
a \left( t \right) & =
A_{\pm} \, t^{1 \pm \nu}
& & (K=0,~ \widetilde \lambda_{*} >0) \label{7.6a} \\
a \left( t \right) & =
\frac{t}{2 \, \nu} \, \left[ \left( \frac{t}{t_{1}} \right)^{\nu}
+ \left( \frac{t_{1}}{t} \right)^{\nu} \right]
& & (K=+1,~ \widetilde \lambda_{*} >0) \label{7.6b} \\
a \left( t \right) & =
\frac{t}{2 \, \nu} \, \left[ \left( \frac{t}{t_{1}} \right)^{\nu}
- \left( \frac{t_{1}}{t} \right)^{\nu} \right]
& & (K=-1,~ \widetilde \lambda_{*} >0) \label{7.6c} \\
a \left( t \right) & =
\frac{t}{\nu} \, \cos \left[ \nu \, \ln \left( t / t_{1} \right) \right]
& & (K=-1,~ \widetilde \lambda_{*} <0) \label{7.6d}.
\end{align}
\end{subequations}
Here $\nu \equiv \sqrt{\lvert \widetilde \lambda_{*} \rvert / 3 \,}>0$ and
$A_{\pm} \equiv b_{0} \, {t_{1}}^{\mp \nu} \, \sqrt{\widetilde g_{*} /
\, \overline{G} \,}$.
\subsubsection{Solutions from the Einstein static Universe}
Another well-known solution of the $b$-equations with $\overline{\Lambda}>0$,
$K=+1$ is the Einstein static Universe, here filled with radiation rather
than dust:
\begin{gather}
b = \left( \frac{3}{2} \, \frac{1}{\, \overline{\Lambda} \,} 
\right)^{1/2},
\quad
\check \rho = 
\frac{1}{8 \pi} \, \frac{\overline{\Lambda}}{\, \overline{G} \,}. \label{7.7}
\end{gather}
For arbitrary $G \left( t \right)$ it generates the solution
\begin{gather}
a \left( t \right) = \sqrt{\frac{3 \, G \left( t \right)}{2
\, \overline{G} \, \overline{\Lambda} \,} \,},
\quad
\rho \left( t \right) = \frac{\overline{G} \, \overline{\Lambda}}
{8 \pi \, G \left( t \right)^{2}},
\quad (K=+1). \label{7.8} 
\end{gather}
In the fixed point regime it describes a scale-free, linearly expanding 
Universe of positive spatial curvature:
\begin{gather}
a \left( t \right) = \left( \frac{3}{2} \, \frac{1}{\, \widetilde \lambda_{*}
\,}
\right)^{1/2} \, t,
\quad
\rho \left( t \right) = \frac{1}{8 \pi} \, \frac{\widetilde \lambda_{*}}
{\, \widetilde g_{*} \,} \, t^{-4}. \label{7.9} 
\end{gather}
This cosmology is quite similar to the attractor solution
 found in \cite{cosmo1} by improving
the Einstein equations. The only difference is that $\xi$
does not get fixed in the present approach.
\subsubsection{Another radiation Universe}
As a last example, we consider the standard spatially flat, radiation
dominated Universe with a positive cosmological constant:
\begin{align}
b \left( \tau \right) & =
\left[ \frac{{\mathcal M} \overline{G}}{2 \, \overline{\Lambda} \,} \,
\biggl \{ \cosh \left( 4 \, \sqrt{\, \overline{\Lambda} / 3 \,} ~ \tau
\right) -1 \biggr \} \right]^{1/4} \label{7.10}.
\end{align}
In the fixed point regime with $\widetilde \lambda_{*}>0$, it generates a 
$K=0$, $w=1/3$ cosmology with the scale factor
\begin{align}
a \left( t \right) & =
\left( \frac{{\mathcal M} \widetilde g_{*}}{4 \, \widetilde \lambda_{*} \,}
\right)^{1/4} \, t \,
\left[ \left( \frac{t}{t_{1}} \right)^{4 \nu} +
\left( \frac{t_{1}}{t} \right)^{4 \nu} - 2 \right]^{1/4} \label{7.11}.
\end{align}
For $\nu < 1$ this cosmology has no initial singularity.
\subsection{The Class \textrm{\textbf{IIIb}}}
In Class IIIb, which constitutes the generic case and is much larger than
Class IIIa, solutions cannot be found by the Weyl-transformation technique.
One has to work directly with the system of equations \eqref{7.1}.
In general it is over-determined and cannot be solved for arbitrary
backgrounds $\left( G \left( t \right), \Lambda \left( t \right) \right)$.
To find some illustrative examples of cosmologies in Class IIIb we confine
ourselves to the power law backgrounds \eqref{5.25} which we employed in 
Class I already. It turns out that \eqref{7.1} possesses the following 
three families of power law solutions with $K=0$, $w \geq -1$,
and ${\mathcal M} \neq 0$.
\subsubsection*{1. First family $\boldsymbol{\Bigl( n \neq \pm 2,~ n \neq 4 /
\!\left( 1+3w \right),~ w>-1,~w \neq -1/3 \Bigr)}$}
This family of solutions is the Class IIIb-counterpart to the Class 
I-solutions in the ``First family'' of Subsection \ref{4.C}.
Its members are labeled by the free constants ${\mathcal M}$ and $C>0$, while
$n$ and $D$ are fixed by the requirement of solubility:
\begin{subequations} \label{7.12}
\begin{align}
a \left( t \right) & =
\left[ \frac{6 \, \left( 1+w \right)^{2}}
{\left( n-2 \right) \, \left( n+3\,nw-4 \right)}
\, {\mathcal M} \, C \right]^{1 / \left( 3+3w \right)} \,
t^{\left( n+2 \right) / \left( 3+3w \right)}, \label{7.12a}\\
\rho \left( t \right) & =
\frac{1}{48 \pi} \frac{\left( n-2 \right) \, \left( n+3\,nw-4 \right)}
{\left( 1+w \right)^{2}} \, \frac{1}{C} \, t^{- \left( n+2 \right)}, \label{7.12b}\\
G \left( t \right) & = C \, t^{n}, \label{7.12c}\\
\Lambda \left( t \right) & =
\frac{n \, \left( 3w-1 \right) \, \left( n+3\, nw -4 \right)}
{12 \,\left( 1+w \right)^{2}}  \,
t^{-2}. \label{7.12d}
\end{align}
\end{subequations}
For $n=0$ this result reduces to the classical Friedmann cosmology with
$K=0$ and $\Lambda =0$. In the general case, the allowed values of the exponent
$n$ are subject to certain restrictions which depend on the sign of 
${\mathcal M}$.
The solution exists and $a \left( t \right)$ is real if $n$, $w$, and
${\mathcal M}$ are such that
\begin{gather}
\left( n-2 \right) \, \left( n + 3 \, nw -4 \right) / {\mathcal M} >0.
\label{7.13}
\end{gather}

For $w=1/3$, say\footnote{Strictly speaking, for this particular equation of 
state the cosmology \eqref{7.12} belongs to Class II rather than IIIb 
because it fulfills both defining conditions, $w=1/3$ and $\Lambda=0$. It is
more convenient to consider it as a special member of the above family, 
however. It can also be obtained as a Weyl-transform by using \eqref{6.21}\
with $G \left( t \right) = C \, t^{n}$, $n \neq 2$, and $t_{1}=0$ (for
$n<2$) or $t_{1} \to \infty$ (for $n>2$).},
this requirement is met by any real $n \neq \pm2$ if ${\mathcal M} >0$, and it
cannot be satisfied at all if ${\mathcal M} <0$. Thus, for an almost arbitrary
exponent $n$, we always obtain a solution with a positive matter energy
density $\rho$.

Taking $w=0$ as a second example, we see that for all exponents $n<2$ and 
$n>4$ we again have ${\mathcal M} >0$ and positive energy density therefore.
Only for $2<n<4$ we need a negative ${\mathcal M}$ and $\rho$.

It is instructive to compare the Class IIIb-cosmology \eqref{7.12} to its
Class I-analog \eqref{5.26} which was computed for the same time dependence
of $G$. We observe that the various functions contain the same powers of $t$
but different prefactors, and those different prefactors lead to different
conditions for the existence of positive and negative energy density solutions,
respectively. In Class I, solutions with ${\mathcal M} >0$ exist only in a 
finite interval of $n$-values; the ${\mathcal M} <0$-solutions are more
abundant and are realized in two infinite bands of exponents. In Class IIIb
the situation is exactly the other way around: The ${\mathcal M} >0$-solutions
are the more abundant ones and obtain for an infinite range of $n$-values.
Solutions with ${\mathcal M} <0$ exist at most within a finite $n$-interval.

These findings provide a further confirmation of the general picture we 
developed earlier. In Class III, contrary to Class I, a $\theta$-tensor is
included which can absorb the energy and momentum carried by $G \left( x 
\right)$ and $\Lambda \left( x \right)$. As a consequence, the solutions
do not need to ``squeeze'' those contributions into $T_{\mu \nu}$ which then
has a chance of describing ordinary matter with positive energy density and 
pressure.
\subsubsection*{2. Second family $\boldsymbol{\Bigl( n=-2,~ w>-1 \Bigr)}$}
If $n=-2$, solutions exist only if $m=+2$ so that both
$G$ and $\Lambda$ are proportional to $t^{-2}$.
The remarkable property of this
cosmology is that its scale factor and energy density are constant,
the Universe does not expand, even though $G$ and $\Lambda$ have a nontrivial
time dependence:
\begin{subequations} \label{7.14}
\begin{align}
a \left( t \right) & = \left[ \frac{1}{4} \, \left( 1+w \right) \,
{\mathcal M} \, C \right]^{1 / \left( 3+3w \right)}
= const, \label{7.14a}\\
\rho \left( t \right) & = \frac{1}{2 \pi} \frac{1}{\left( 1+w \right)} \,
\frac{1}{C}
= const, \label{7.14b}\\
G \left( t \right) & = C \, t^{-2}, \label{7.14c} \\
\Lambda \left( t \right) & = \frac{\left( 3w-1 \right)}{\left( 1+w \right)} \,
t^{-2}. \label{7.14d}
\end{align}
\end{subequations}
(Since $\dot a=0$ here, the $ii$-component of Einstein's equation has been
checked explicitly in the present case.) The cosmology \eqref{7.14} is similar
to \eqref{5.29} but contrary to the latter it has a positive energy density
which confirms the general picture.
\subsubsection{Third family $\boldsymbol{\Bigl( n=-m \neq -2,~w=-1 \Bigr)}$}
This family describes expanding or contracting Universes whose matter energy
density remains constant:
\begin{subequations} \label{7.15}
\begin{align}
a \left( t \right) & =
A \, t^{n/2}, ¸\label{7.15a} \\
\rho \left( t \right) & = 
\frac{{\mathcal M}}{8 \pi}, \label{7.15b} \\
G \left( t \right) & =
C \, t^{n}, \label{7.15c} \\
\Lambda \left( t \right) & =
- {\mathcal M} C \, t^{n}. \label{7.15d}
\end{align}
\end{subequations}
The overall scale $A$ is arbitrary and ${\mathcal M}$ can have either sign.
%
%
\section{Summary and Conclusion} \label{8}
In this paper we discussed the general framework describing the gravitational dynamics in presence of position-dependent ``constants'' $G \left( x \right)$ 
and $\Lambda \left( x \right)$ which result from RG-improving the 
Einstein-Hilbert action. The $x$-dependence of $G$ and $\Lambda$ is governed
by a RG trajectory on a truncated theory space, together with a cutoff
identification $k=k \left( x \right)$ relating spacetime points to RG scales.
The improvement is effected by the replacement 
$G \rightarrow G \left( x \right)$, $\Lambda \rightarrow \Lambda \left( x 
\right)$ in the Lagrangian density. The resulting formalism has a certain
similarity with Brans-Dicke theory, but there are also crucial differences.
In particular, $G \left( x \right)$ and $\Lambda \left( x \right)$
are externally prescribed background fields in the present case since the
RG equations admit no simple (local) Lagrangian description. We derived the
modified Einstein field equations, and we showed that their consistency
imposes certain conditions upon the scalar fields $G \left( x \right)$ 
and $\Lambda \left( x \right)$. The main property both theories have in common
is that those scalar fields carry energy and momentum and contribute to the
curvature of spacetime therefore. As for the energy-momentum tensor 
pertaining to $G$ and $\Lambda$, either theory contains the piece
$\Delta T_{\mu \nu}$ which results from varying $\sqrt{-g} \, R / G \left(
x \right)$ with respect to the metric. In addition, conventional Brans-Dicke
theory contains a term ${\mathcal T}^{\omega}_{\mu \nu}$ which is a consequence
of the (almost) standard kinetic term $\propto \omega \, \left( D \phi
\right)^{2}$ which Brans and Dicke postulated for the field 
$\phi \equiv 1/G$. This term has no immediate analog within the present 
framework since the RG flow is not described in a Lagrangian setting.
Therefore, to be as general as possible, we included an a priori arbitrary 
tensor $\theta_{\mu \nu}$ in the modified Einstein equation which, 
together with
$\Delta T_{\mu \nu}$, is to describe the 4-momentum residing
in the $x$-dependence of $G$ and $\Lambda$. The form of $\theta_{\mu \nu}$
is severely constrained by the consistency condition. In the extreme case
of $\Lambda \equiv 0$ and a traceless energy-momentum tensor of the matter
fields (``Class II''), for instance, $\theta_{\mu \nu}$ gets uniquely fixed. 
We found that
$\theta_{\mu \nu}=\theta_{\mu \nu}^{\text{\scriptsize BD}}$ where
$\theta_{\mu \nu}^{\text{\scriptsize BD}}$ equals the Brans-Dicke tensor
${\mathcal T}^{\omega}_{\mu \nu}$ for the exceptional value $\omega = -3/2$.
From the point of view of ordinary Brans-Dicke theory, $\omega = -3/2$ 
amounts to the singular limit where the Klein-Gordon equation for $\phi$ 
decouples and no longer determines or constrains $G \left( x \right)$. The
reason is that precisely for $\omega = -3/2$ the kinetic term 
$\propto \omega \, \left( D \phi \right)^{2}$ can be absorbed into the 
$\sqrt{-g} \, R$-term by a Weyl-rescaling of the metric. This mechanism allows
us to treat $G \left( x \right)$ as an externally prescribed field then.
It also provides us with a remarkably simple and efficient method for
solving the in general rather complicated modified Einstein equations 
with $\theta_{\mu \nu}=\theta_{\mu \nu}^{\text{\scriptsize BD}}$. If 
$T_{\mu}^{~\mu} =0$ and either $\Lambda \equiv 0$ (``Class II'') or
$\Lambda \left( x  \right) \propto 1 / G \left( x  \right)$
(``Class IIIa''), all solutions can be obtained by Weyl-transforming solutions
of the corresponding Einstein equations with constant $G$ and $\Lambda$,
which are solved much more easily, of course. The classes II and IIIa include
applications which are of particular physical interest. In ``small'' systems
such as black holes, say, the cosmological constant does not play a central
role typically and may be neglected so that we are in Class II, and in all 
situations where the underlying RG trajectory is close to a fixed point one has
$\Lambda \left( x  \right) \propto 1 / G \left( x  \right)$ and we are in
Class IIIa. The condition on the matter system, the tracelessness of 
$T_{\mu \nu}$, is satisfied most trivially in the vacuum, but clearly one may
also think of classical radiation or a (quantum) conformal field theory.

Once we have solved the RG equations for $G \left( k \right)$ and
$\Lambda \left( k \right)$, and have converted the $k$-dependence to a
$x$-dependence, various strategies for exploiting this information in a 
dynamical context suggest themselves. In the present approach we replaced
$G \rightarrow G \left( x \right)$, $\Lambda \rightarrow \Lambda \left( x 
\right)$ in the Lagrangian density; alternatively one could,
for instance, first derive the standard Einstein equation from the classical
action in the usual way, and then replace 
$G \rightarrow G \left( x \right)$, $\Lambda \rightarrow \Lambda \left( x 
\right)$ at the level of the equation of motion.
In general the field equations obtained by the two methods are different.
Those obtained by varying the improved action functional contain terms
involving derivates of $G$ (the tensor $\Delta T_{\mu \nu}$) which could
never arise by improving the equation of motion. As a result, in the former
approach, there is energy and momentum associated to the variation of $G$ in time
and space, while this is not the case in the latter.

Thus at first sight it might seem that the two approaches lead to quite
different physical predictions so that at least one of them should be wrong.
In this paper we demonstrated that this is actually not the case and that,
provided both methods are applicable, they can very well lead to identical or
at least qualitatively similar results. However, matching the two approaches
is not straightforward. In particular, it requires either a re-interpretation
of the ``matter'' energy-momentum tensor $T_{\mu \nu}$, or the inclusion of an
appropriate $\theta$-tensor.

In order to clarify these issues we chose Robertson-Walker cosmology as a 
first application because it is a typical and at the same time technically
simple and transparent example. For the same reason we employed the time 
dependence of $G$ and $\Lambda$ arising from a RG trajectory near a fixed 
point in most of our examples\footnote{We also performed analogous calculations
in the ``perturbative regime'' where one expands in powers of 
$k / m_{\text{\scriptsize Pl}}$ or $t_{\text{\scriptsize Pl}} / t$,
respectively, so that one can see the transition from the classical to the
quantum domain. Since the results are quite lengthy and not particularly 
instructive we do not display them here.}.
For the choice $\theta_{\mu \nu} =0$ we found the following rather surprising
result: If one interprets $T_{\mu \nu} + \Delta T_{\mu \nu}$ rather than
$T_{\mu \nu}$ alone as the matter energy-momentum tensor, the spatially
flat fixed point cosmologies obtained by both approaches coincide
\textit{exactly} if $w = 1/3$ and approximately for other values of $w$.
Also in many other cosmologies we observed the same phenomenon. In Section
\ref{5} we discussed and explained it in detail, and in Section \ref{6} and
\ref{7} we showed that the ``contamination'' of the matter energy-momentum 
tensor by contributions due to the $x$-dependence of $G$ and $\Lambda$
can be avoided if one allows for an non-zero $\theta$-tensor. In our
examples we adapted the choice 
$\theta_{\mu \nu}=\theta_{\mu \nu}^{\text{\scriptsize BD}}$, both because
this tensor is the one used in ordinary Brans-Dicke theory and because
of its uniqueness property mentioned above. In the classes II and IIIa,
thanks to this choice, we were able to obtain strikingly simple closed-form
solutions to the quite complicated differential equations governing the
cosmological evolution, some of them valid for arbitrary $G \left( t \right)$
even.

The upshot of our general discussion and the analysis of the
cosmological examples is that in a proper application of the
improved-Lagrangian approach one should include a $S_{\theta}$-term into
the total action. We saw that there is a dual motivation for it: it makes the
modified Einstein equation consistent, and at the same time it brings the
improved-action approach closer to the improved-equation method.
Even if the analogy is not complete we can think of $S_{\theta}$
as an analogue of the nonpolynomial potential term
$\propto \Phi^{4} \, \ln (\Phi)$ or the nonlocal kinetic term
$\propto \Phi \, Z \left( \sqrt{- \partial^{2}\,} \right) \, \partial^{2} \Phi$
discussed in the Introduction. While not contained in the truncation ansatz,
the RG reasoning suggests that these terms should be present in $\Gamma$.

In conclusion we can say that in this paper we have gained 
a physical understanding of how
to interpret the results from the improved action-approach and how to relate
them to those obtained earlier by improving the field equations. The new
approach has a much wider range of applicability than the older one, and as we
now understand how to handle it properly it will be possible to apply it
to situations where the improvement of the field equations makes no sense.
The most important example of this kind are vacuum spacetimes in absence
of a cosmological constant, the Schwarzschild black hole, for instance.
They satisfy $G_{\mu \nu}=0$, and clearly this equation is completely
``blind'' to a possible $x$-dependence of Newton's constant. It will be 
interesting therefore to investigate the quantum properties of black holes
within the framework developed in this paper. We shall come back to this
point elsewhere \cite{holger2}.\\[24pt]
Acknowledgment: We would like to thank A.~Bonanno for helpful discussions.
\\
%
%
\appendix
\begin{flushleft}
{\Large \textbf{Appendix}}
\end{flushleft}
\section{Improving the field equations} \label{appendix A}
In this appendix we collect some of the results of ref.\ \cite{cosmo1}
which are needed in the main text. In \cite{cosmo1} the approach of RG
improving field equations (as opposed to solutions or actions) has been
applied to cosmology. The starting point is the standard Einstein equation
without additional terms, $G_{\mu \nu} = - \Lambda \, g_{\mu \nu}
+ 8 \pi G \, T_{\mu \nu}$. The RG improvement consists of replacing
$G \rightarrow G \left( t \right)$, $\Lambda \rightarrow \Lambda \left( t 
\right)$ in this equation. Specializing for a Robertson-Walker metric, one
obtains the following system of coupled equations:
\begin{subequations} \label{A.1}
\begin{gather}
\left( \frac{\dot a}{a} \right)^{2} + \frac{K}{a^{2}} =
\frac{1}{3} \, \Lambda + \frac{8 \pi}{3} \, G \, \rho \label{A.1a} \\
\dot \rho + 3 \, \left( 1+w \right) \, \left( \frac{\dot a}{a} \right) \,
\rho =0 \label{A.1b}\\
\dot \Lambda + 8 \pi \, \rho \, \dot G =0. \label{A.1c}
\end{gather}
\end{subequations}
Eq.\ \eqref{A.1a} has the form of the standard Friedmann equation with a 
time-dependent $G$ and $\Lambda$ inserted, eq.\ \eqref{A.1b} is the usual
continuity equation, and eq.\ \eqref{A.1c} is the consistency condition 
resulting from the integrability condition 
$D^{\mu} \left[ - \Lambda \, g_{\mu \nu} + 8 \pi G \, T_{\mu \nu} \right] =0$.

With $G \left( k \right)$ and $\Lambda \left( k \right)$ in the fixed point
regime and the cutoff identification $k = \xi / t$ the time dependence
of $G$ and $\Lambda$ is given by \eqref{1.50}. For this time dependence, and
a fixed value of the density parameter ${\mathcal M}$, the system
\eqref{A.1} has the following unique solution:
\begin{subequations} \label{A.50}
\begin{align}
a \left( t \right) & = 
\left[ \left( \frac{3}{8} \right)^{2} \left( 1+w \right)^{4} \, {\mathcal M} 
\, g_{*} \lambda_{*} \right]^{1\, / \, \left( 3+3w \right)}
\, t^{4\, / \, \left( 3+3w \right)} \label{A.50a}\\
\rho \left( t \right) & = 
\frac{8}{9 \pi} \, \frac{1}{\left( 1+w \right)^{4}} \, 
\frac{1}{g_{*} \lambda_{*}} \, \frac{1}{t^{4}}  \label{A.50b}\\
G \left( t \right) & = 
\frac{3}{8} \left( 1+w \right)^{2} \, g_{*} \lambda_{*}\,  t^{2} \label{A.50c}
\\
\Lambda \left( t \right) & = 
\frac{8}{3} \frac{1}{\left( 1+w \right)^{2}} \, \frac{1}{t^{2}} \label{A.50d}.
\end{align}
\end{subequations}
The integrability of \eqref{A.1} fixes the constant $\xi$ according to
\begin{gather}
\xi^{2}
= \frac{8}{3 \, \left( 1+w \right)^{2}} \, \frac{1}{\lambda_{*}}. \label{A.51}
\end{gather}
For a detailed discussion of the cosmology \eqref{A.50} we refer to
\cite{cosmo1,kalligas}. Further solutions of the system \eqref{A.1}, in
particular in the perturbative regime of the renormalization group
(expansion in powers of $k / m_{\text{\scriptsize Pl}}$) can be found in
\cite{cosmo1}.
%
%
\section{Solving the consistency condition \eqref{2.26}} 
\label{appendix B}
In this appendix we prove that $\vartheta_{\mu \nu}^{\text{\scriptsize BD}}$
of eq.\ \eqref{2.27} is the unique tensor which satisfies the consistency
condition \eqref{2.26} for \textit{all} functions $\psi$ and which vanishes
for $\psi = const$. (For special $\psi$'s further solutions might exist.)

To start with, we look for the most general solution $\vartheta_{\mu \nu}$,
constructed from $\psi$ and its derivatives, which contains no more than two
derivatives. We make an ansatz
\begin{align}
\begin{split}
\vartheta_{\mu \nu} & = 
A \left( \psi \right) \, D_{\mu} \psi \, D_{\nu} \psi
+ B \left( \psi \right) \, g_{\mu \nu}
\, \left( D \psi \right)^{2}
 \\
& \phantom{{==}} 
+ C \left( \psi \right) \, D_{\mu} D_{\nu} \psi
+ D \left( \psi \right) \, g_{\mu \nu} \, D^{2} \psi
+ g_{\mu \nu} \, E \left( \psi \right)
\end{split}
\label{B.1}
\end{align}
with $A,B,C,\cdots$ arbitrary functions of $\psi$. Upon inserting 
\eqref{B.1} into \eqref{2.26} combinations of those functions and their
derivatives multiply various field monomials such as
$D_{\mu} \psi \, D^{\mu} \psi \, D^{\nu} \psi$, 
$D^{2} \psi \, D^{\nu} \psi$, etc.; since $\psi$ is assumed arbitrary,
these monomials are linearly independent and so their prefactors must
vanish separately. This leads to a system of differential equations for
$A,B,C,\cdots$ whose general solution can be found easily. Inserting
it into \eqref{B.1} we obtain
\begin{align}
\vartheta_{\mu \nu} & =  - \frac{3}{2} 
\, \left[ D_{\mu} \psi \, D_{\nu} \psi
- \frac{1}{2} \, g_{\mu \nu} \, \left( D \psi \right)^{2} \right]
+ g_{\mu \nu} \, E (0) \, e^{\psi}
\label{B.2}
\end{align}
The only free constant of integration is $E (0)$. As $\vartheta_{\mu \nu}$ must
vanish for $\psi = const$ we are forced to set $E (0) =0$, in which case
$\vartheta_{\mu \nu}$ becomes equal to 
$\vartheta_{\mu \nu}^{\text{\scriptsize BD}}$.

In a second step, we write the most general solution as
\begin{align}
\vartheta_{\mu \nu} & = 
\vartheta_{\mu \nu}^{\text{\scriptsize BD}} + \widehat \vartheta_{\mu \nu}.
\label{B.3}
\end{align}
Since $\vartheta_{\mu \nu}^{\text{\scriptsize BD}}$ is a special solution to 
the inhomogeneous equation \eqref{2.26}, $\widehat \vartheta_{\mu \nu}$
is the general solution of the homogeneous equation
\begin{align}
D^{\mu} \widehat \vartheta_{\mu \nu}
+ \Bigl( \widehat \vartheta_{\mu \nu} - \tfrac{1}{2} \, g_{\mu \nu} \,
\widehat \vartheta_{\alpha}^{~\alpha} \Bigr) \, D^{\mu} \psi =0.
\label{B.4}
\end{align}
It simplifies when we rewrite it in terms of
$\widehat \theta_{\mu \nu} \equiv  \widehat 
\vartheta_{\mu \nu} / \left( 8 \pi G \right)$ which enters the analogous 
decomposition 
$\theta_{\mu \nu} = 
\theta_{\mu \nu}^{\text{\scriptsize BD}} + \widehat \theta_{\mu \nu}$:
\begin{align}
D^{\mu} \widehat \theta_{\mu \nu} & =
\tfrac{1}{2} \, \widehat \theta_{\alpha}^{~\alpha}  \, D_{\nu} \psi.
\label{B.5}
\end{align}
We assume that $\widehat \theta_{\mu \nu}$ is generated by the action
$\widehat S_{\theta} \bigl[ g_{\mu \nu}, \psi \bigr]$:
\begin{align}
\widehat \theta^{\mu \nu} & =
\frac{2}{\sqrt{-g \,}\,} \, 
\frac{\delta \widehat S_{\theta}}{\delta g_{\mu \nu}}.
\label{B.6}
\end{align}
The action $\widehat S_{\theta}$ is invariant under general coordinate
transformations if $g_{\mu \nu}$ and $\psi$ transform as a tensor and a 
scalar, respectively. Therefore, by the same argument which lead to 
\eqref{2.19},
\begin{align}
D^{\mu} \widehat \theta_{\mu \nu} & =
\frac{1}{\sqrt{-g \,}\,} \, 
\frac{\delta \widehat S_{\theta}}{\delta \psi} \, D_{\nu} \psi.
\label{B.7}
\end{align}
Using \eqref{B.7} in \eqref{B.5} for $D_{\mu} \psi \neq 0$ the problem boils
down to finding the general solution of
\begin{align}
\left[ g_{\mu \nu} \, \frac{\delta}{\delta g_{\mu \nu}}
- \frac{\delta}{\delta \psi} \right] \, 
\widehat S_{\theta} [g, \psi] = 0.
\label{B.8}
\end{align}
Eq.\ \eqref{B.8} is the infinitesimal form of the invariance condition
\begin{align}
\widehat S_{\theta} \bigl[ e^{\alpha} \, g_{\mu \nu}, ~\psi-\alpha \bigr]
& = \widehat S_{\theta} \bigl[ g_{\mu \nu}, \psi \bigr],
\label{B.9}
\end{align}
where $\alpha \left( x \right)$ is an arbitrary parameter. Eq.\ \eqref{B.9}
tells us that $\widehat S_{\theta}$ is not a functional of 
$g_{\mu \nu}$ and $\psi$ separately but only of the combination
$e^{\psi} \, g_{\mu \nu} \equiv \gamma_{\mu \nu}$.
Writing $\widehat S_{\theta} [g, \psi] = F \bigl[ e^{\psi} \, g \bigr]
\equiv F [ \gamma]$, the corresponding energy-momentum tensor reads
\begin{align}
\widehat \theta^{\mu \nu} [g,\psi] & =
e^{3 \psi} \, \left(
\frac{2}{\sqrt{- \gamma \,}\,} \, 
\frac{\delta F [\gamma]}{\delta \gamma_{\mu \nu}} \right)
\, \bigl[ \gamma_{\alpha \beta} = e^{\psi} \, g_{\alpha \beta} \bigr].
\label{B.10}
\end{align}

At this point the condition that $\theta_{\mu \nu}$ must vanish for
$\psi = const$ plays an important role. Since
$\theta_{\mu \nu}^{\text{\scriptsize BD}}$ does have this property it follows
that $\widehat\theta_{\mu \nu}$ must vanish separately for constant $\psi$.
However, for $F [\gamma]$ an arbitrary functional, the tensor \eqref{B.10}
does not in general vanish for $\psi = const$. Thus we see that tensors
of the form \eqref{B.10} are not admissible, except when $F [\gamma] =0$,
$ \widehat\theta_{\mu \nu}=0$. This concludes our proof that the unique,
identical solution of \eqref{2.26} is the Brans-Dicke-type tensor
$\vartheta_{\mu \nu}^{\text{\scriptsize BD}}$.
%
%
\section{Eliminating the $\boldsymbol{ii}$-components} \label{C}
In this appendix we prove that, under the condition $\dot a \neq 0$,
every set of functions $a \left( t \right)$, $\rho \left( t \right)$,
$p \left( t \right)$ which satisfies the $00$-component of the modified
Einstein equation, the consistency condition, and the continuity equation 
\eqref{4.5} automatically satisfies the $ii$-components of the modified
Einstein equation as well.

The demonstration proceeds as follows: Defining the tensor
$T_{\mu \nu}^{\prime}$ by
\begin{align}
\begin{split}
8 \pi \overline{G} \, T_{\mu \nu}^{\prime} & \equiv
- g_{\mu \nu} \, \Lambda + 8 \pi G \, \left(
T_{\mu \nu} + \Delta T_{\mu \nu} + \theta_{\mu \nu} \right) \\
& = - g_{\mu \nu} \, \Lambda + 8 \pi G \, T_{\mu \nu}
+ \Delta t_{\mu \nu} + \vartheta_{\mu \nu} 
\end{split} \label{C.1}
\end{align}
with an arbitrary, fixed constant $\overline{G}$, Einstein's equation
\eqref{2.10} assumes the standard form
\begin{align}
G_{\mu \nu} & = 8 \pi \overline{G} \, \, T_{\mu \nu}^{\prime} \label{C.2}.
\end{align}
For symmetry reasons, $T_{\mu}^{\prime \, \nu}$ has the structure
$T_{\mu}^{\prime \, \nu} = \text{diag} \left[ - \rho^{\prime},
p^{\prime},p^{\prime},p^{\prime} \right]$ 
with $\rho^{\prime}$ and $p^{\prime}$ depending on $t$ only. In 
$T_{\mu \nu}^{\prime}$-language, since $\overline{G}$ is constant, the
consistency condition \eqref{2.20} is equivalent to the statement that
$T_{\mu \nu}^{\prime}$ has vanishing covariant divergence, i.\,e.\ that
$D_{\mu} T^{\prime \mu \nu} = 0$, or
\begin{gather}
\dot \rho^{\prime} + 3 \, H \, \left( \rho^{\prime} + p^{\prime} \right)
=0. \label{C.3}
\end{gather}
Moreover, the $00$- and $ii$-components of \eqref{C.2}
assume the same form as in standard cosmology without a cosmological constant,
albeit with a complicated equation of state $p^{\prime}=p^{\prime}
\left( \rho^{\prime} \right)$. The $00$-component is
\begin{subequations} \label{C.4}
\begin{gather}
H^{2} + \frac{K}{a^{2}} = \frac{8 \pi}{3} \, \overline{G} \, \rho^{\prime}
\label{C.4a},
\intertext{and the $ii$-component reads}
H^{2} + 2 \, \left( \frac{\ddot a}{a} \right) + \frac{K}{a^{2}} =
- 8 \pi \overline{G} \, p^{\prime}. \label{C.4b}
\end{gather}
\end{subequations}
If functions $\rho^{\prime} \left( t \right)$, $p^{\prime} \left( t \right)$,
$a \left( t \right)$ satisfy \eqref{C.3} and \eqref{C.4a}, they also
satisfy \eqref{C.4b}. 

This can be seen by differentiating \eqref{C.4a} with respect to $t$,
\begin{gather}
2 \, \left( \frac{\ddot a}{a} \right) \, H
- 2\, H^{3} - 2 \, \frac{K}{a^{2}} \, H =
\frac{8 \pi}{3} \, {\overline{G}} \, \dot \rho^{\prime}, \nonumber
\intertext{and substituting \eqref{C.3} for $\dot \rho^{\prime}$:}
2 \, \left( \frac{\ddot a}{a} \right) \, H
- 2\, H^{3} - 2 \, \frac{K}{a^{2}} \, H =
- 8 \pi \, \overline{G} \, H\, \left( \rho^{\prime} + p^{\prime} \right).
\nonumber
\intertext{This equation implies}
2 \, H^{2} + 2 \, \frac{K}{a^{2}} =
2 \, \left( \frac{\ddot a}{a} \right) + 8 \pi \, 
\overline{G} \, \left( \rho^{\prime} + p^{\prime} \right), 
\label{C.5}
\end{gather}
provided $H \neq 0$. Inserting \eqref{C.5} into \eqref{C.4a} according to
\begin{align*}
8 \pi \, \overline{G} \rho^{\prime} & =
\left[2 \, H^{2} + 2 \, \frac{K}{a^{2}} \right]
+ \left[H^{2} + \, \frac{K}{a^{2}} \right]\\
& = 2 \, \left( \frac{\ddot a}{a} \right) + 8 \pi \, 
\overline{G} \, \left( \rho^{\prime} + p^{\prime} \right)
+ \left[ H^{2} + \, \frac{K}{a^{2}} \right],
\end{align*}
leads precisely to the $ii$-component \eqref{C.4b}, which is thus
seen to be a consequence of the $00$-component and the continuity equation
if the scale factor $a \left( t \right)$ is not constant, i.\,e.\ if 
$H \neq 0$.

Thus, rather than \eqref{C.4a} and \eqref{C.4b}, one may use the 
$00$-component \eqref{C.4a} and the conservation law for 
$T_{\mu \nu}^{\prime}$, \eqref{C.3}, as independent equations.
Returning now to the original formulation without the primed quantities, it is
clear that the integrability condition, $D_{\mu}
T^{\prime \mu \nu}=0$, is nothing but the off-shell 
consistency condition \eqref{2.20} or, more explicitly, \eqref{4.14}. (In 
deriving it we used indeed that the divergence of \eqref{C.1} must be
zero and assumed that the ordinary continuity equation 
$D_{\mu} T^{\mu \nu}=0$ is valid.)
This shows that if a cosmology with $a \neq const$
satisfies the $00$-component of Einstein's equation, eq.\ \eqref{4.9a}, the
consistency condition \eqref{4.14} and the ordinary continuity condition
\eqref{4.5}, then, for any equation of state, it also satisfies the
$ii$-component of Einstein's equation.
%
%

\begin{thebibliography}{99}
\bibitem{bradi}
C.~Brans and R.H.~Dicke, 
Phys.\ Rev.\ 124 (1961) 925.
%
\bibitem{bertodm}
O.~Bertolami et al., 
Phys.\ Lett.\ B 311 (1993)27; \\
O.~Bertolami and J.~Garcia-Bellido, 
Int.\ J.\ Mod.\ Phys.\ D 5 (1996) 363 .
%
\bibitem{avact}
C.~Wetterich, 
Phys.\ Lett.\ B 301 (1993) 90.
%
\bibitem{avactrev}
For a review see: 
J.~Berges, N.~Tetradis and C.~Wetterich, 
Phys.\ Rep.\ 363 (2002) 223;
C.~Wetterich, 
Int.\ J.\ Mod.\ Phys.\ A 16 (2001) 1951.
%
\bibitem{ym}
M.~Reuter and C.~Wetterich, 
Nucl.\ Phys.\ B 417 (1994) 181,
Nucl.\ Phys.\ B 427 (1994) 291, 
Nucl.\ Phys.\ B 391 (1993) 147, 
Nucl.\ Phys.\ B 408 (1993) 91; \\
M.~Reuter, 
Phys.\ Rev. D 53 (1996) 4430, 
Mod.\ Phys.\ Lett.\ A 12 (1997) 2777.
%
\bibitem{mr}
M.~Reuter, 
Phys.\ Rev.\ D 57 (1998) 971 and hep-th/9605030.
%
\bibitem{percadou}
D.~Dou and R.~Percacci, 
Class.\ Quant.\ Grav.\ 15 (1998) 3449.
%
\bibitem{carfora}
M.~Carfora and K.~Piotrkowska, 
Phys.\ Rev.\ D 52 (1995) 4393; \\
T.~Buchert and M.~Carfora, 
gr-qc/0101070; \\
T.~Buchert and M.~Carfora,
Phys.\ Rev.\ Lett.\ 90 (2003) 031101.
%
\bibitem{bagber}
For a review see: 
C.~Bagnuls and C.~Bervillier, 
Phys.\ Rep.\ 348 (2001) 91; \\
T.R.~Morris, 
Prog.\ Theor.\ Phys.\ Suppl.\ 131 (1998) 395.
%
\bibitem{cwnonloc}
C.~Wetterich, 
Gen.\ Rel.\ Grav.\ 30 (1998) 159.
%
\bibitem{oliver1}
O.~Lauscher and M.~Reuter, 
Phys.\ Rev.\ D 65 (2001) 025013 and hep-th/0108040.
%
\bibitem{frank1}
M.~Reuter and F.~Saueressig, 
Phys.\ Rev.\ D 65 (2002) 065016 and hep-th/0110054.
%
\bibitem{oliver2}
O.~Lauscher and M.~Reuter, 
Class.\ Quant.\ Grav.\ 19 (2002) 483 and hep-th/0110021; 
Phys.\ Rev.\ D 66 (2002) 025026 and hep-th/0205062; \\
Int.\ J.\ Mod.\ Phys.\ A 17 (2002) 993 and hep-th/0112089.
%
\bibitem{percacciperini}
R.~Percacci and D.~Perini, 
Phys.\ Rev.\ D 67 (2003) 081503;
Phys.\ Rev.\ D 68 (2003) 044018 ; 
D.~Perini, 
hep-th/0305053.
%
\bibitem{frank2}
M.~Reuter and F.~Saueressig, 
Phys.\ Rev.\ D 66 (2002) 125001 and hep-th/0206145.
%
\bibitem{venez}
G.~Veneziano, 
Mod.\ Phys.\ Lett.\ A 4 (1989) 695; \\
T.~Taylor and G.~Veneziano, 
Phys.\ Lett.\ B 228 (1990) 210.
%
\bibitem{nelson}
B.~L.~Nelson, P.~Panangaden,
Phys.\ Rev.\ D 25 (1982) 1019.
%
\bibitem{cosmo1}
A.~Bonanno and M.~Reuter,
Phys.\ Rev.\ D 65 (2002) 043508 and hep-th/0106133.
%
\bibitem{cosmo2}
A.~Bonanno and M.~Reuter,
Phys.\ Lett.\ B 527 (2002) 9 and astro-ph/0106468; \\
Int.\ J.\ Mod.\ Phys.\ D, in press, and astro-ph/0210472; \\
E.~Bentivegna, A.~Bonanno and M.~Reuter,
astro-ph/0303150.
%
\bibitem{esposito}
A.~Bonanno, G.~Esposito and C.~Rubano,
hep-th/0303154.
%
\bibitem{bh}
A.~Bonanno and M.~Reuter, 
Phys.\ Rev.\ D 62 (2000) 043008 and hep-th/0002196;
Phys.\ Rev.\ D 60 (1999) 084011 and gr-qc/9811026.
%
\bibitem{kalligas}
D.~Kalligas, P.~Wesson and C.W.F.~Everitt, 
Gen.\ Rel.\ Grav.\ 24 (1992)351; \\
for a similar cosmology in a Brans-Dicke framework see: \\
O.~Bertolami and P.J.~Martins,
Phys.\ Rev.\ D 61 (2000) 064007.
%
\bibitem{souma}
W.~Souma,
Prog.\ Theor.\ Phys.\ 102 (1999) 181.
%
\bibitem{wein}
S.~Weinberg 
in \textit{General Relativity, an Einstein Centenary Survey},
S.W.~Hawking and W.~Israel (Eds.), 
Cambridge University Press (1979);
S.~Weinberg,
hep-th/9702027.
%
\bibitem{max}
P.~Forg\'acs and M.~Niedermaier, 
hep-th/0207028; \\
M.~Niedermaier,
JHEP 0212 (2002) 066;
Nucl.\ Phys.\ B 673 (2003) 131.
%
\bibitem{bertocosmo}
O.~Bertolami,
Nuovo Cim.\ B 93 (1986) 36.
%
\bibitem{sola}
I.L.~Shapiro and J.~Sol\`a,
Phys.\ Lett.\ B 475 (2000) 236;
JHEP 02 (2002) 006; \\
I.L.~Shapiro, J.~Sol\`a, C.~Espa\~na-Bonet and P.~Ruiz-Lapuente,
Phys.\ Lett.\ B 574 (2003) 149.
%
\bibitem{mrcwcosmo}
M.~Reuter and C.~Wetterich, 
Phys.\ Lett.\ B 188 (1987) 38.
%
\bibitem{cwquint}
C.~Wetterich,
Nucl.\ Phys.\ B 302 (1988) 668; 
Astron.\ Astrophys.\ 301 (1995) 321.
%
\bibitem{quint}
P.~J.~Peebles and B.~Ratra,
Astrophys.\ J.\ Lett.\ 325 (1988) L 17; \\
B.~Ratra and P.~J.~Peebles,
Phys.\ Rev.\ D 37 (1988) 3406.
%
\bibitem{faraoui}
For a review and a comprehensive list of references see:
V.~Faraoni, E.~Gunzig and P.~Nardone,
Fund.\ Cosmic Phys.\ 20 (1999) 121.
%
\bibitem{maeda}
Y.~Fujii and K.~Maeda,
\textit{The scalar-tensor theory of gravitation}, 
Cambridge University Press (2002).
%
\bibitem{blago}
M.~Blagojevi\'c, 
\textit{Gravitation and Gauge Symmetries},
IOP Publishing, Bristol (2002).
%
\bibitem{holger2}
M.~Reuter and H.~Weyer,
in preparation
%
\end{thebibliography}

\end{document}